\documentclass[pra,twocolumn,floatfix,superscriptaddress,amsmath,aps,longbibliography]{revtex4-1}
\usepackage{graphicx}
\usepackage{color}
\usepackage{amsmath}
\usepackage[colorlinks,plainpages=false,linkcolor=blue,urlcolor=blue,citecolor=blue,pdfpagemode=UseNone,pdfstartview=FitBH]{hyperref}

\newcommand{\crbr}{CrBr$_3$}

\begin{document}

\title{Thermal Evolution of Dirac Magnons in the Honeycomb Ferromagnet CrBr$_3$}

\author{S.~E.~Nikitin}\email[Corresponding author: ]{Stanislav.Nikitin@psi.ch}
\affiliation{Quantum Criticality and Dynamics, Paul Scherrer Institute, 
CH-5232 Villigen-PSI, Switzerland}

\author{B.~F{\aa}k}
\affiliation{Institut Laue-Langevin, 71 avenue des Martyrs, CS 20156, F-38042 
Grenoble Cedex 9, France}

\author{K.~W.~Kr\"{a}mer}
\affiliation{Department of Chemistry, Biochemistry and Pharmacy, University 
of Bern, Freiestrasse 3, CH-3012 Bern, Switzerland}

\author{T.~Fennell}
\affiliation{Laboratory for Neutron Scattering and Imaging, Paul Scherrer 
Institute, CH-5232 Villigen-PSI, Switzerland}

\author{B.~Normand}
\affiliation{Laboratory for Theoretical and Computational Physics, Paul 
Scherrer Institute, CH-5232 Villigen-PSI, Switzerland}
\affiliation{Institute of Physics, Ecole Polytechnique F\'ed\'erale 
de Lausanne (EPFL), CH-1015 Lausanne, Switzerland}

\author{A. M. L\"auchli}
\affiliation{Laboratory for Theoretical and Computational Physics, Paul 
Scherrer Institute, CH-5232 Villigen-PSI, Switzerland}
\affiliation{Institute of Physics, Ecole Polytechnique F\'ed\'erale 
de Lausanne (EPFL), CH-1015 Lausanne, Switzerland}

\author{Ch.~R\"{u}egg}
\affiliation{Quantum Criticality and Dynamics, Paul Scherrer Institute, 
CH-5232 Villigen-PSI, Switzerland}
\affiliation{Institute of Physics, Ecole Polytechnique F\'ed\'erale 
de Lausanne (EPFL), CH-1015 Lausanne, Switzerland}
\affiliation{Institute for Quantum Electronics, ETH Z\"urich, 
CH-8093 H\"onggerberg, Switzerland}
\affiliation{Department of Quantum Matter Physics, University of Geneva, 
CH-1211 Geneva, Switzerland}

\begin{abstract}
CrBr$_3$ is an excellent realization of the two-dimensional honeycomb 
ferromagnet, which offers a bosonic equivalent of graphene with Dirac
magnons and topological character. We perform inelastic neutron scattering 
(INS) measurements using state-of-the-art instrumentation to update 
50-year-old data, thereby enabling a definitive comparison both with recent 
experimental claims of a significant gap at the Dirac point and with 
theoretical predictions for thermal magnon renormalization. We demonstrate 
that CrBr$_3$ has next-neighbor $J_2$ and $J_3$ interactions approximately 
5\% of $J_1$, an ideal Dirac magnon dispersion at the K point, and the 
associated signature of isospin winding. The magnon lifetime and the thermal 
band renormalization show the universal $T^2$ evolution expected from an 
interacting spin-wave treatment, but the measured dispersion lacks the 
predicted van Hove features, highlighting the need for a deeper 
theoretical analysis. 
\end{abstract}

\maketitle

\begin{figure}[t]
\includegraphics[width=1\columnwidth]{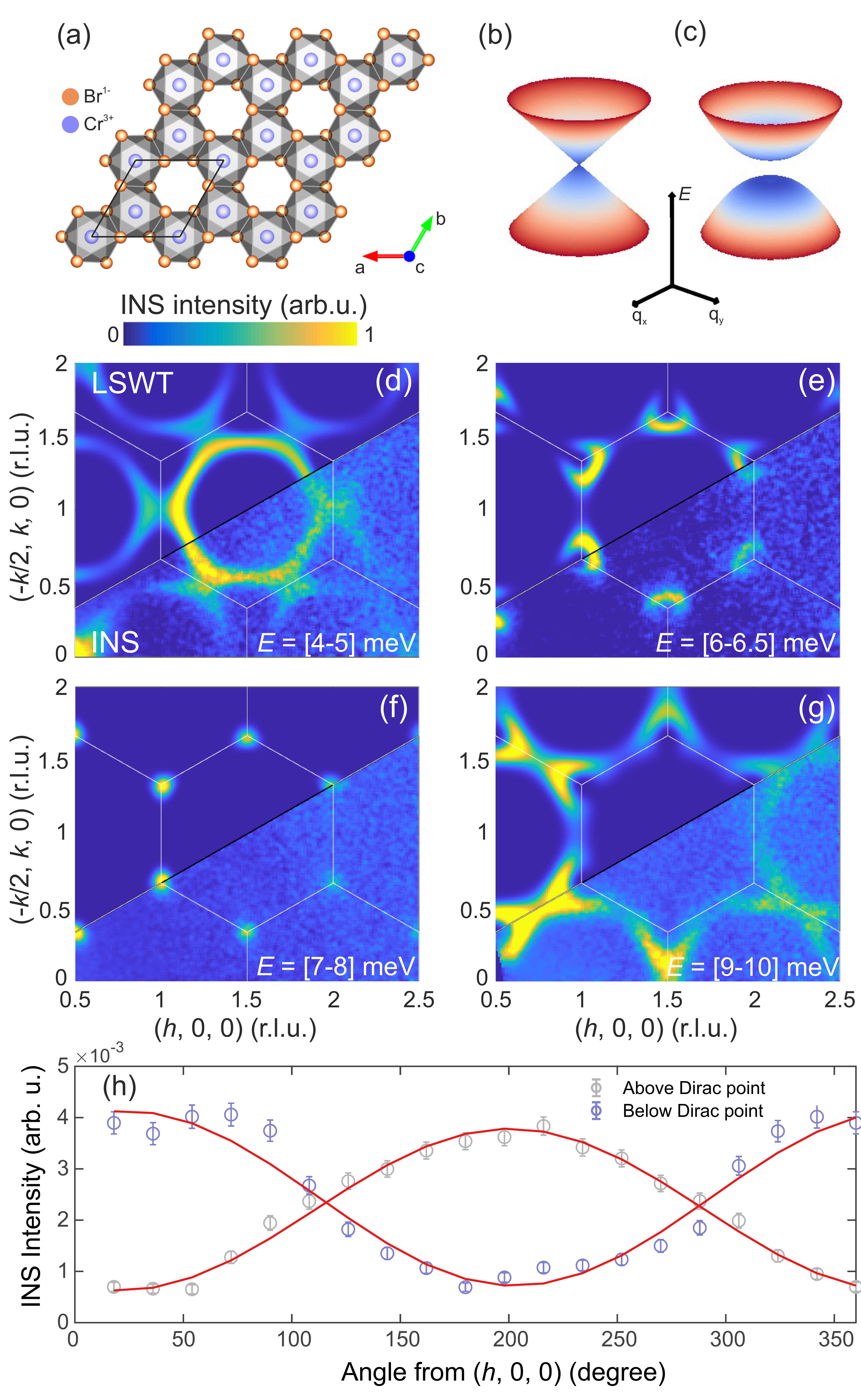}
\caption{
(a)~Honeycomb layer of \crbr. The Cr$^{3+}$ ions (blue) host $S = 3/2$ spins 
with FM interactions. (b,c)~Schematic spin-wave spectra in the vicinity of 
the K points. When inversion symmetry is preserved, $\omega_q \propto q$ and 
the dispersion forms a Dirac cone (b); otherwise a gap opens to form separate 
acoustic and optical magnon branches (c).
(d-g)~Scattered intensity obtained by integrating the PANTHER $E_{\mathrm{i}} = 
30$~meV dataset over four constant-energy windows (indicated) and compared 
with linear spin-wave theory (LSWT); all panels have the same intensity 
scale. White lines indicate the boundaries of the crystallographic BZ. A 
$\mathbf{Q}$-independent background was subtracted from each spectrum to 
aid visual comparison.
(h) Intensity obtained by winding around the K point, showing a cosinusoidal 
modulation with inverted phase for energies above and below the Dirac point.}
\label{fig1}
\end{figure}

Graphene, the original two-dimensional (2D) material, is a single layer of 
carbon atoms with strong covalent bonds forming a honeycomb lattice, and 
some of its exceptional physical properties~\cite{novoselov2004electric, 
novoselov2005two, geim2009graphene} are a consequence of its band-structure 
topology, which allows the electrons to behave as massless quasiparticles 
described by the Dirac equation. The same band structure is realized for 
bosonic quasiparticles in systems such as a 2D ferromagnet (FM) on the 
honeycomb lattice \cite{pershoguba2018dirac} shown in Fig.~\ref{fig1}(a), 
which has magnon excitations that also exhibit Dirac cones at the K points of 
the Brillouin zone (BZ), as represented in Fig.~\ref{fig1}(b). The topology 
of magnon band structures has became a matter of active theoretical 
\cite{li2017dirac, pershoguba2018dirac, mcclarty2021topological, 
mook2021interaction} and experimental \cite{yao2018, chen2018topological,
yuan2020dirac,elliot2021order,cai2021topological,scheie22} research due 
to possible applications in spintronic devices \cite{chumak2015magnon,
wang2018topological,pirro2021advances}. Inelastic neutron scattering 
(INS) provides direct access to the magnon dispersion, and the spectra 
of a number of honeycomb FMs have been measured in their low-temperature 
regimes ($T \ll\ T_c$, the temperature of magnetic order) 
\cite{chen2018topological,elliot2021order,chen2021massless}.

Promising materials for these studies are the family of chromium trihalides, 
Cr$X_3$ ($X = $ Cl \cite{morosin64}, Br \cite{samuelsen1971spin}, I 
\cite{mcguire15}), in which the honeycomb layers [Fig.~\ref{fig1}(a)] have 
identical stacking, but $T_c$, the size, and even the sign of the interlayer 
magnetic interaction all vary with $X$ \cite{wang2011electronic}. 
Measurements on CrCl$_3$ \cite{chen2021massless,do2022} indicate a Dirac-cone 
magnon dispersion [Fig.~\ref{fig1}(b)], but in CrI$_3$ a gap is reported 
\cite{chen2018topological} at the K point, creating acoustic and optical magnon 
modes [Fig.~\ref{fig1}(c)] whose anticrossing is thought to be a consequence 
of strong next-neighbor Dzyaloshinskii-Moriya (DM) interactions. \crbr\ was for 
50 years considered as a textbook example of FM magnons, with no indication for 
a band splitting \cite{samuelsen1971spin, yelon1971renormalization}, but the 
recent report of a large, DM-induced anticrossing \cite{cai2021topological} 
similar to CrI$_3$ has created controversy. Theoretical calculations based on 
a Dirac-cone spectrum have predicted a very specific $T^2$ form for the 
temperature-induced renormalization of the magnon dispersion and line width 
\cite{pershoguba2018dirac}, and cited the old INS results as verification, 
but systematic measurements of the thermal evolution of the magnon spectrum 
remain absent. 

In this Letter we perform a comprehensive study of the temperature-induced 
renormalization of the magnon self-energy in \crbr\ using modern neutron 
spectrometers. We first use low-temperature INS data to refine the magnetic 
spin Hamiltonian and find weak next-neighbor interactions. We prove that the 
magnon dispersion has Dirac cones, the recent report to the contrary apparently 
being an artifact of the data treatment, and we demonstrate near-ideal 
cosinusoidal intensity winding around the K points. Working at temperatures 
up to 40~K, we find considerable downward renormalization of the magnon 
dispersion and growing line widths, whose $T^2$ form we characterize to 
high accuracy, but the variation of these terms across the BZ is not well 
captured by the available theory. Our results set the experimental standard 
for temperature-induced modification of the spin dynamics in a honeycomb 
ferromagnet. 

\begin{figure}[t]
\includegraphics[width=1\columnwidth]{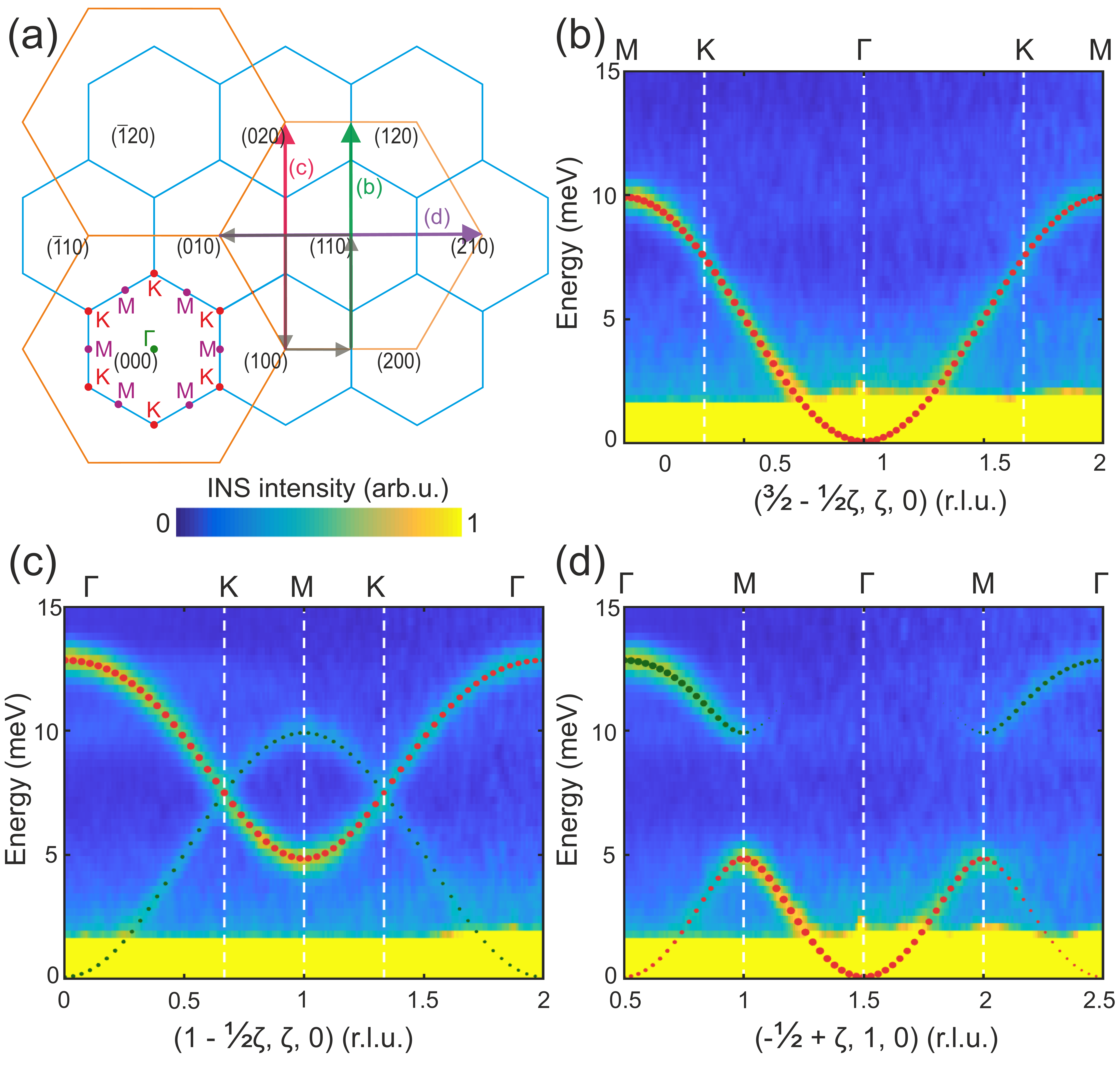}
\caption{Low-temperature magnon dispersion measured along several 
high-symmetry directions.
(a)~Reciprocal-space representation of the honeycomb lattice in the 
$(h \, k \, 0)$ plane showing the crystallographic BZ in blue and the 
unfolded zone in orange. Green, red, and violet arrows indicate the paths 
for the spectra shown respectively in panels (b), (c), and (d); the gray 
arrow indicates the complete path used in Fig.~\ref{fig3}.
(b-d)~Spin-wave spectra collected using PANTHER with $E_{\mathrm{i}} = 30$~meV 
at $T = 1.7$~K. The data were integrated by $\pm 0.03$ r.l.u.~along the 
orthogonal in-plane direction and $\pm 5$ r.l.u.~in the out-of-plane $l$ 
direction. Red and green points show respectively the dispersions of modes 
1 and 2 modelled by LSWT and their sizes represent the calculated intensities.}
\label{fig2}
\end{figure}

\textit{Experiment.} A 1.5~g single crystal of \crbr\ was grown 
by slow sublimation in a temperature gradient under vacuum, as detailed 
in Sec.~S1 of the Supplementary Materials (SM)~\cite{SI}. Its high quality 
was confirmed by single-crystal neutron diffraction, from which we determined 
the lattice parameters at 1.7 K as $a = b = 6.31$~\AA\ and $c = 18.34$~\AA, 
and confirmed the BiI$_3$-type structure with space group R${\overline 3}$. 
We conducted two INS experiments, using the time-of-flight (TOF) spectrometer 
PANTHER at the Institut Laue-Langevin \cite{panther,ILLdoi} and the 
triple-axis spectrometer (TAS) EIGER at the Paul Scherrer Institute 
\cite{stuhr2017}. In both experiments the sample was oriented in the $(h \, 
k \, 0)$ scattering plane. On PANTHER we collected data at $T = 1.7$, 20, 30, 
and 40~K, each with two incident neutron energies, $E_{\mathrm{i}} = 15 $ and 
30~meV, and performed TOF data reduction and analysis using the software 
\textsc{MANTID} \cite{Mantid} and \textsc{HORACE} \cite{Horace}. On EIGER we 
used the fixed-$k_{\mathrm{f}}$ mode and worked at eight different temperatures 
from 1.5 to 40~K. Calculations of the low-temperature magnon dispersion and 
intensity, which we used to fit the spin Hamiltonian, were performed using 
the \textsc{SpinW} package \cite{toth15}.
 
\begin{figure*}[t]
\includegraphics[width=\textwidth]{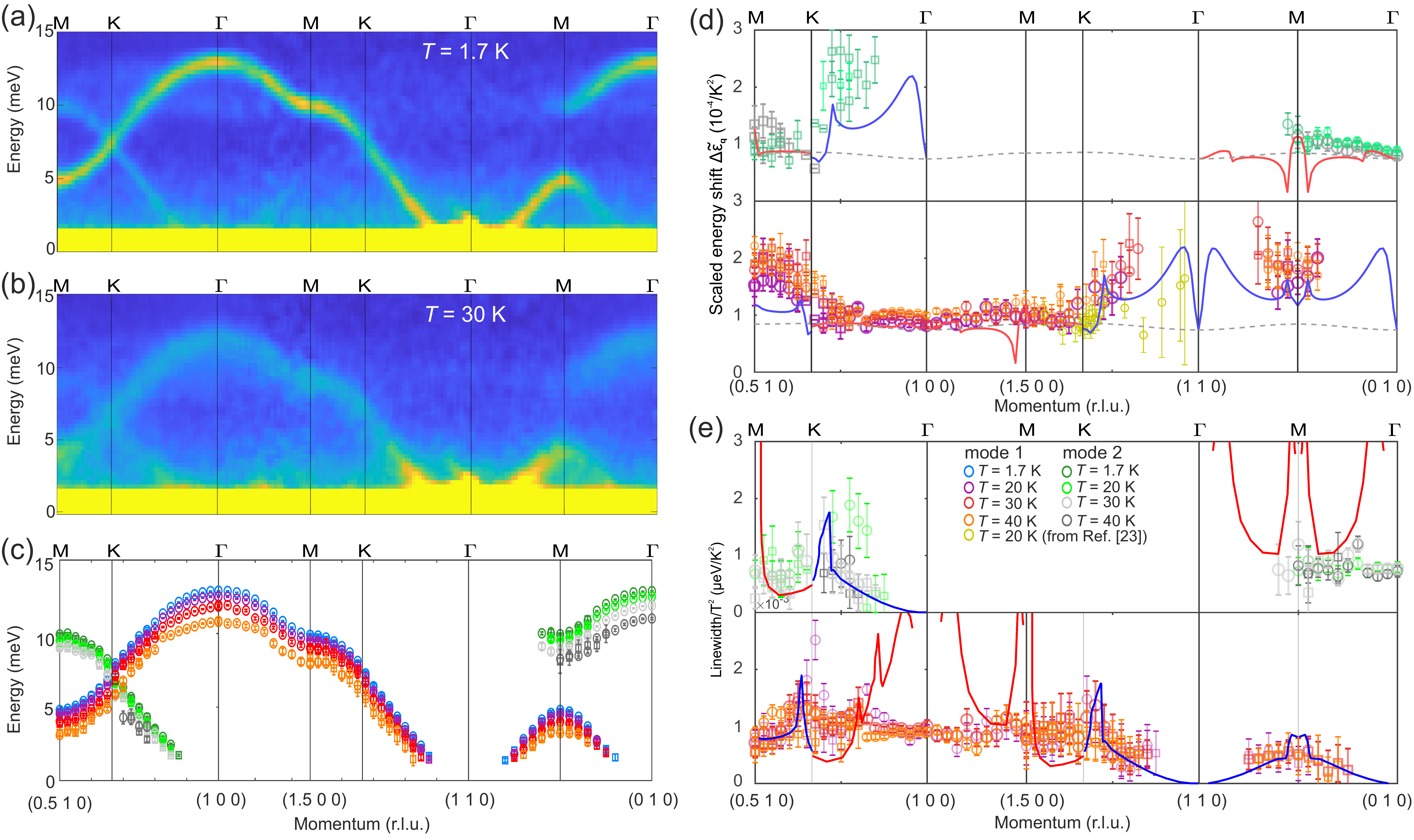}
\caption{Thermal magnon renormalization in \crbr. 
(a,b) Magnon spectra for the high-symmetry directions taken from the PANTHER 
$E_{\mathrm{i}} = 30$~meV dataset at $T = 1.7$~K (a) and $T = 30$~K (b). The data 
were integrated over $\pm 0.015$ \AA$^{-1}$ in-plane and $\pm 2$ \AA$^{-1}$ for 
$l$. (c) Magnon branches 1 and 2 extracted for the four experimental 
temperatures; circles were taken from the $E_{\mathrm{i}} = 30$~meV dataset 
and squares from $E_{\mathrm{i}} = 15$~meV. The mode intensities vanish in 
some regions of the unfolded BZ. 
(d) Temperature-induced renormalization of the measured magnon dispersions 
for modes 1 (upper) and 2 (lower panel), shown in the reduced form of 
Eq.~\eqref{Eq:deltaE}. Solid lines show the real part of the self-energy, 
${\rm Re} \, \Sigma (\mathbf{q})$, obtained by adapting the analytical 
framework of Ref.~\cite{pershoguba2018dirac}, in which calculations are 
performed for the upper (red) and lower (blue) bands in the crystallographic 
BZ; dashed lines show the Hartree term, $\Sigma_{1} (\mathbf{q})$.
(e) Reduced thermal renormalization of the measured magnon line widths 
for modes 1 (upper) and 2 (lower panel). Solid lines show $- {\rm Im} \, 
\Sigma (\mathbf{q})$ obtained following Ref.~\cite{pershoguba2018dirac}.}
\label{fig3}
\end{figure*}

\textit{Low-temperature spectra.}
We begin with the spectra collected on PANTHER at $T = 1.7$~K, a temperature 
much smaller than $T_c = 32$~K \cite{alyoshin1997rf,cai2021topological} and 
thus fully representative of the ground-state properties. Figures 
\ref{fig1}(d-g) show constant-energy cuts at four different parts of the 
magnon spectral function and Figs.~\ref{fig2}(b-d) show momentum-energy cuts 
for several high-symmetry paths in the BZ [Fig.~\ref{fig2}(a)]. We also used 
the vertical detector coverage to confirm dispersionless behavior in the
out-of-plane direction, as shown in Sec.~S2A of the SM \cite{SI}. Focusing 
first on the two M-K-$\Gamma$ paths in Figs.~\ref{fig2}(b) and \ref{fig2}(c), 
both spectra exhibit a sharp, continuous, and resolution-limited magnon mode 
with a band width of approximately 10~meV, a parabolic dispersion around 
$\Gamma$, and different intensities in the two zones shown. Figures 
\ref{fig2}(c) and \ref{fig2}(d) show the second magnon branch in the 
crystallographic BZ dispersing from 5 to 13~meV, although with zero 
intensity in Fig.~\ref{fig2}(b), and we refer to the two branches as 
modes 1 and 2. Here we label all high-symmetry points according to the 
crystallographic BZ, but stress that the modulation of the scattered 
intensity follows the unfolded zone shown in Fig.~\ref{fig2}(a), leading 
to the intensity variations between BZs in Figs.~\ref{fig1} and \ref{fig2}.

\begin{figure}[t]
\includegraphics[width=1\columnwidth]{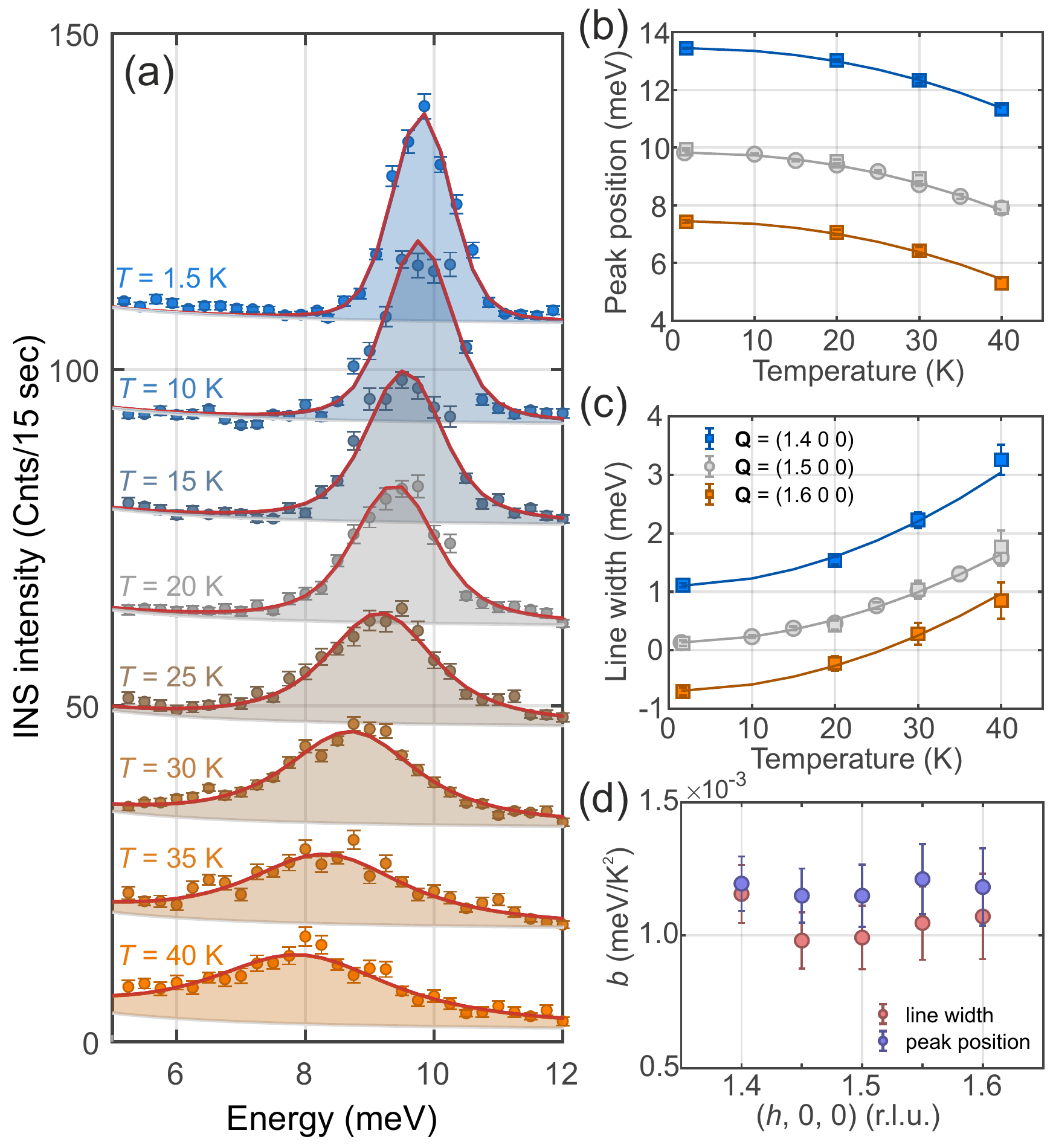}
\caption{Thermal magnon renormalization at the M point.
(a) Constant-$\mathbf{Q}$ cuts measured on EIGER at the M point, $\mathbf{Q}
 = (1.5~0~0)$. Spectra for the different temperatures are offset vertically 
by $+15$ units for clarity. Red lines show complete fits to the data composed 
of a Gaussian magnon contribution (shaded areas) and a temperature-independent 
background. 
(b,c) Dependence of the peak position (b) and line width (c) on temperature; 
solid lines show quadratic fits. Squares and circles correspond respectively 
to results obtained by fitting the PANTHER and EIGER datasets.
(d) $\mathbf{Q}$-dependence of the quadratic fitting coefficients, $b$, 
through the M point.}
\label{fig4}
\end{figure}

Our first key result is the unambiguous demonstration of the data in 
Figs.~\ref{fig2}(b) and \ref{fig2}(c) that the magnon bands have a Dirac 
dispersion through the K point, with no detectable splitting into acoustic and 
optical modes. It is important to contrast this conclusion with the recent 
INS study of Ref.~\cite{cai2021topological}, which reported a large band 
splitting at the K point. In Sec.~S2B of the SM \cite{SI} we demonstrate that 
the reported splitting is not intrinsic to \crbr, but is rather an artifact 
arising from the large integration width applied in the analysis of the TOF 
dataset \cite{do2022}. 

Thus we conclude that the low-temperature magnon dispersion in \crbr\ has 
an ideal Dirac-cone nature with the Dirac point at $7.5 \pm 0.1$ meV 
[Fig.~\ref{fig1}(f)]. This is fully consistent with the inversion symmetry 
of the nearest-neighbor bond and the conventional $g$-factor values, both of 
which exclude significant DM effects. It is also consistent with all of the 
early INS results \cite{samuelsen1971spin,yelon1971renormalization}, as we 
show in Sec.~S2C of the SM \cite{SI}. The Dirac cone in the 2D honeycomb FM 
was also used as a test case for the theoretical prediction 
\cite{shivam2017neutron} of a cosinusoidal intensity modulation arising from 
the isospin winding of near-nodal quasiparticles. This fingerprint has been 
observed recently in the honeycomb material CoTiO$_3$ \cite{elliot2021order} 
and in elemental Gd \cite{scheie22}, and our results for the intensity 
distribution around the K point, shown in Fig.~\ref{fig1}(h) and detailed in 
Sec.~S2D of the SM \cite{SI}, constitute its cleanest observation to date.

Next we use our low-temperature INS spectra to refine the spin Hamiltonian. 
Based on the lack of evidence for DM interactions in Fig.~\ref{fig2}, but 
the very accurate measurement of a tiny spin gap at the $\Gamma$ point by 
ferromagnetic resonance (FMR) \cite{dillon1962ferromagnetic,alyoshin1997rf}, 
we consider a Heisenberg model with single-ion anisotropy,
\begin{equation}
\mathcal{H} = \sum_{\langle i,j \rangle} J_{ij} \mathbf{S}_i \cdot \mathbf{S}_j
 + D \sum_i (S_i^z)^2.
\label{Eq:Hamiltonian}
\end{equation}
Here $\mathbf{S}_i$ is a $S = 3/2$ spin operator, $J_{ij}$ are isotropic 
superexchange interactions between different Cr-ion pairs, and $D$ is the 
single-ion term. We determine the energies of the two magnon modes at 139 
$\mathbf{Q}$ points by fitting the corresponding constant-$\mathbf{Q}$ 
cuts to two resolution-convolved Lorentz functions. We then use this dataset 
to fit the magnetic interactions in \crbr\ by working within LSWT, as 
implemented in \textsc{SpinW}. We find the most accurate description of 
the observed spectra using three in-plane interactions and a very weak 
easy-axis anisotropy, as detailed in Sec.~S3 of the SM~\cite{SI}. The 
optimal parameters we obtain for Eq.~\eqref{Eq:Hamiltonian} are $J_1 = - 
1.485(15)$, $J_2 = - 0.077(13)$, $J_3 = 0.068(12)$, and $D = - 0.028(7)$~meV. 
Although we cannot detect the spin gap created by such a small anisotropy, 
we include the gap deduced from FMR in our fit. The excellent agreement 
between the observed and calculated INS spectra, both in dispersion and 
intensity distribution, is clear in Figs.~\ref{fig1}(d-g) and \ref{fig2}(b-d).

\textit{Spin dynamics at finite temperature.} Turning to thermal effects, 
Figs.~\ref{fig3}(a) and \ref{fig3}(b) show two representative spectra 
collected respectively at $T = 1.7$ and 30~K. Increasing $T$ clearly broadens 
the magnons and causes a downward energy shift, which decreases their band 
width. To quantify both effects, and their dependence on $\mathbf{Q}$, we 
used PANTHER to measure the spectral function at $T = 20$, 30, and 40~K 
over several BZs. We made multiple constant-$\mathbf{Q}$ cuts covering four 
high-symmetry directions and fitted each peak with a Lorentzian broadening, 
convolved with the experimental resolution, to extract the positions and 
widths of the two magnon modes at each $T$ and $\mathbf{Q}$ point. Figure 
\ref{fig3}(c) summarizes the mode positions obtained at all four temperatures. 

To visualize the effect of temperature on the magnon bands, we compute the 
normalized dispersion shift \cite{yelon1971renormalization}
\begin{equation}
\Delta {\tilde \varepsilon}_{\mathbf{q}}(T) = \frac{\varepsilon_{\mathbf{q}}(0)
 - \varepsilon_{\mathbf{q}}(T)}{\varepsilon_{\mathbf{q}}(0)T^2}, 
\label{Eq:deltaE}
\end{equation}
where $\varepsilon_{\mathbf{q}}(0)$ denotes the dispersion measured at base 
temperature and $\varepsilon_{\mathbf{q}}(T)$ the corresponding finite-$T$ 
result. In the interacting SWT analysis of Ref.~\cite{pershoguba2018dirac}, 
the $T$-induced dispersion renormalization consists of a real Hartree 
term, $\Sigma_1 (\mathbf{q})$, with a weak $\mathbf{q}$-dependence caused 
only by $J_2$, and a ``sunset'' term, $\Sigma_2 (\mathbf{q})$. Because 
both are expected to show a $T^2$ form \cite{bloch30,pershoguba2018dirac}, 
we have included this factor in Eq.~\eqref{Eq:deltaE}.

The symbols in Fig.~\ref{fig3}(d) show the dispersion renormalization along 
the high-symmetry paths. The data for different temperatures collapse rather 
well to a single curve for both modes over the majority of the BZ, and we find 
that no change to the assumed $T^2$ form improves this collapse. To interpret 
this result, we have adapted the calculations of Ref.~\cite{pershoguba2018dirac}
to include the $J_2$ and $J_3$ terms, and present the details of this 
adaptation in Sec.~S4 of the SM \cite{SI}. We observe that $\Delta {\tilde 
\varepsilon}_{\mathbf{q}}(T)$ for the upper branch is described largely by the 
Hartree term alone, with the $\Sigma_2(\mathbf{q})$ contribution becoming 
sizeable only below the Dirac point. 
 
Similarly, Fig.~\ref{fig3}(e) demonstrates the analogous $T^2$ data 
reduction for the magnon line width. Again the experimental results for all 
temperatures collapse rather well, within their own uncertainties, to a single 
line. In this case, $J_2$ and $J_3$ have a qualitative role in removing
line-width divergences that appear at the $\Gamma$ and M points due to the 
perfect nesting of the nearest-neighbor bands \cite{pershoguba2018dirac}. 
However, even with these terms, the interacting SWT analysis predicts that 
both the line width and the band renormalization [Fig.~\ref{fig3}(d)] should 
show multiple sharp peaks across the BZ, these ``van Hove'' features 
reflecting the underlying bare magnon bands \cite{pershoguba2018dirac}, 
whereas our data do not support their presence. 

A striking example is the difference between our data and the adapted SWT 
treatment around the M point, where the analysis predicts that both the energy 
and width of the 10 meV should show a sharp cusp, which is shifted slightly 
from M due to $J_2$ and $J_3$ [red lines in Figs.~\ref{fig3}(d) and 
\ref{fig3}(e)]. To analyze the thermal renormalization in a fully quantitative 
manner, we used EIGER to measure the spectrum at the M point $\mathbf{Q} = 
(1.5~0~0)$ for multiple temperatures up to 40~K, as shown in Fig.~\ref{fig4}(a).
Figures \ref{fig4}(b) and \ref{fig4}(c) show respectively the dependences on 
$T$ of the magnon energy and line width extracted from both EIGER and PANTHER 
datasets. When fitted to the form $a + b T^\alpha$, the M-point data yield 
$\alpha_{\mathrm{Energy}} = 2.25(15)$ and $\alpha_{\mathrm{Width}} = 1.95(14)$, in 
good agreement with the expected value, $\alpha = 2$. The same fitting at 
several $\mathbf{Q}$ points around ($1.5~0~0$) also yields quadratic forms 
for both quantities [Figs.~\ref{fig4}(b,c)], while the prefactors $b$ that 
we extract show no appreciable changes with $\mathbf{Q}$ in Fig.~\ref{fig4}(d), 
quite in contrast to interacting SWT. 

\textit{Discussion.} 
Our studies of thermal renormalization verify an ideal $T^2$ form, in fact 
above as well as below $T_{\mathrm{c}}$, as we have demonstrated in particular 
detail at the M point (Fig.~\ref{fig4}). The origin of this behavior lies in 
the 2D nature of \crbr\ and the quadratic dispersion at the band minimum, 
where thermally activated magnons cause the interaction effects responsible 
for band renormalization \cite{bloch30}. By a $T^2$ data reduction across the 
whole BZ, we find that the finite-$T$ magnon bands we have measured at high 
$\mathbf{q}$-resolution do not show features at the characteristic wave vectors 
found in a SWT analysis. This indicates that the $S = 3/2$ honeycomb FM is 
subject to complex renormalization effects, arising from the combination of 
quantum and thermal fluctuations in the restricted phase space, whose accurate 
calculation calls for a more advanced (self-consistent and perhaps constrained) 
spin-wave treatment or for an unbiased numerical analysis by state-of-the-art 
quantum Monte Carlo \cite{becker18,shao22} or matrix-product techniques 
\cite{zauner18,ponsioen22}. 

To conclude, we have applied modern neutron spectrometry and data analysis 
to the layered honeycomb $S = 3/2$ ferromagnet \crbr. At the band minimum 
we demonstrate quadratically dispersing magnons with a spin gap far below 
our base temperature. At the K point we demonstrate a near-perfect 
Dirac-cone dispersion with no discernible gapping, and we show that its 
topological consequences are reflected in the intensity winding. We obtain 
an accurate fit of the weak next-neighbor Heisenberg interactions, which 
remove the perfect honeycomb band nesting. At finite temperatures, the magnon 
renormalization obeys the expected $T^2$ form to very high accuracy. However, 
its dependence on the wave vector is not well reproduced at low order in 
spin-wave theory, indicating a need for more systematic calculations of mutual 
quantum and thermal renormalization effects in low-dimensional magnetism. 

{\it Acknowledgments.} We acknowledge financial support from the Swiss 
National Science Foundation, the European Research Council grant Hyper 
Quantum Criticality (HyperQC), and the European Union Horizon 2020 research 
and innovation program under Marie Sk\l{}odowska-Curie Grant No.~884104.
We thank the Institut Laue-Langevin and the Paul Scherrer Institute for 
the allocation of neutron beam-time.

{\it Note added.} During the completion of this manuscript we became 
aware that the data-analysis problem affecting the conclusions of 
Ref.~\cite{cai2021topological}, which we analyze in Sec.~S2B of the SM 
\cite{SI}, has been demonstrated simultaneously for the sister compound 
CrCl$_3$ in Ref.~\cite{do2022}.

\bibliographystyle{apsrev4-1}
\bibliography{main}

\begin{thebibliography}{41}%
\makeatletter
\providecommand \@ifxundefined [1]{%
 \@ifx{#1\undefined}
}%
\providecommand \@ifnum [1]{%
 \ifnum #1\expandafter \@firstoftwo
 \else \expandafter \@secondoftwo
 \fi
}%
\providecommand \@ifx [1]{%
 \ifx #1\expandafter \@firstoftwo
 \else \expandafter \@secondoftwo
 \fi
}%
\providecommand \natexlab [1]{#1}%
\providecommand \enquote  [1]{``#1''}%
\providecommand \bibnamefont  [1]{#1}%
\providecommand \bibfnamefont [1]{#1}%
\providecommand \citenamefont [1]{#1}%
\providecommand \href@noop [0]{\@secondoftwo}%
\providecommand \href [0]{\begingroup \@sanitize@url \@href}%
\providecommand \@href[1]{\@@startlink{#1}\@@href}%
\providecommand \@@href[1]{\endgroup#1\@@endlink}%
\providecommand \@sanitize@url [0]{\catcode `\\12\catcode `\$12\catcode
  `\&12\catcode `\#12\catcode `\^12\catcode `\_12\catcode `\%12\relax}%
\providecommand \@@startlink[1]{}%
\providecommand \@@endlink[0]{}%
\providecommand \url  [0]{\begingroup\@sanitize@url \@url }%
\providecommand \@url [1]{\endgroup\@href {#1}{\urlprefix }}%
\providecommand \urlprefix  [0]{URL }%
\providecommand \Eprint [0]{\href }%
\providecommand \doibase [0]{http://dx.doi.org/}%
\providecommand \selectlanguage [0]{\@gobble}%
\providecommand \bibinfo  [0]{\@secondoftwo}%
\providecommand \bibfield  [0]{\@secondoftwo}%
\providecommand \translation [1]{[#1]}%
\providecommand \BibitemOpen [0]{}%
\providecommand \bibitemStop [0]{}%
\providecommand \bibitemNoStop [0]{.\EOS\space}%
\providecommand \EOS [0]{\spacefactor3000\relax}%
\providecommand \BibitemShut  [1]{\csname bibitem#1\endcsname}%
\let\auto@bib@innerbib\@empty
\bibitem [{\citenamefont {Novoselov}\ \emph {et~al.}(2004)\citenamefont
  {Novoselov}, \citenamefont {Geim}, \citenamefont {Morozov}, \citenamefont
  {Jiang}, \citenamefont {Zhang}, \citenamefont {Dubonos}, \citenamefont
  {Grigorieva},\ and\ \citenamefont {Firsov}}]{novoselov2004electric}%
  \BibitemOpen
  \bibfield  {author} {\bibinfo {author} {\bibfnamefont {K.~S.}\ \bibnamefont
  {Novoselov}}, \bibinfo {author} {\bibfnamefont {A.~K.}\ \bibnamefont {Geim}},
  \bibinfo {author} {\bibfnamefont {S.~V.}\ \bibnamefont {Morozov}}, \bibinfo
  {author} {\bibfnamefont {D.-e.}\ \bibnamefont {Jiang}}, \bibinfo {author}
  {\bibfnamefont {Y.}~\bibnamefont {Zhang}}, \bibinfo {author} {\bibfnamefont
  {S.~V.}\ \bibnamefont {Dubonos}}, \bibinfo {author} {\bibfnamefont {I.~V.}\
  \bibnamefont {Grigorieva}}, \ and\ \bibinfo {author} {\bibfnamefont {A.~A.}\
  \bibnamefont {Firsov}},\ }\href {\doibase 10.1126/science.1102896} {\bibfield
   {journal} {\bibinfo  {journal} {Science}\ }\textbf {\bibinfo {volume}
  {306}},\ \bibinfo {pages} {666} (\bibinfo {year} {2004})}\BibitemShut
  {NoStop}%
\bibitem [{\citenamefont {Novoselov}\ \emph {et~al.}(2005)\citenamefont
  {Novoselov}, \citenamefont {Geim}, \citenamefont {Morozov}, \citenamefont
  {Jiang}, \citenamefont {Katsnelson}, \citenamefont {Grigorieva},
  \citenamefont {Dubonos},\ and\ \citenamefont {Firsov}}]{novoselov2005two}%
  \BibitemOpen
  \bibfield  {author} {\bibinfo {author} {\bibfnamefont {K.~S.}\ \bibnamefont
  {Novoselov}}, \bibinfo {author} {\bibfnamefont {A.~K.}\ \bibnamefont {Geim}},
  \bibinfo {author} {\bibfnamefont {S.~V.}\ \bibnamefont {Morozov}}, \bibinfo
  {author} {\bibfnamefont {D.}~\bibnamefont {Jiang}}, \bibinfo {author}
  {\bibfnamefont {M.~I.}\ \bibnamefont {Katsnelson}}, \bibinfo {author}
  {\bibfnamefont {I.}~\bibnamefont {Grigorieva}}, \bibinfo {author}
  {\bibfnamefont {S.}~\bibnamefont {Dubonos}}, \ and\ \bibinfo {author}
  {\bibfnamefont {A.}~\bibnamefont {Firsov}},\ }\href {\doibase
  10.1038/nature04233} {\bibfield  {journal} {\bibinfo  {journal} {Nature}\
  }\textbf {\bibinfo {volume} {438}},\ \bibinfo {pages} {197} (\bibinfo {year}
  {2005})}\BibitemShut {NoStop}%
\bibitem [{\citenamefont {Geim}(2009)}]{geim2009graphene}%
  \BibitemOpen
  \bibfield  {author} {\bibinfo {author} {\bibfnamefont {A.~K.}\ \bibnamefont
  {Geim}},\ }\href {\doibase 10.1126/science.1158877} {\bibfield  {journal}
  {\bibinfo  {journal} {Science}\ }\textbf {\bibinfo {volume} {324}},\ \bibinfo
  {pages} {1530} (\bibinfo {year} {2009})}\BibitemShut {NoStop}%
\bibitem [{\citenamefont {Pershoguba}\ \emph {et~al.}(2018)\citenamefont
  {Pershoguba}, \citenamefont {Banerjee}, \citenamefont {Lashley},
  \citenamefont {Park}, \citenamefont {{\AA}gren}, \citenamefont {Aeppli},\
  and\ \citenamefont {Balatsky}}]{pershoguba2018dirac}%
  \BibitemOpen
  \bibfield  {author} {\bibinfo {author} {\bibfnamefont {S.~S.}\ \bibnamefont
  {Pershoguba}}, \bibinfo {author} {\bibfnamefont {S.}~\bibnamefont
  {Banerjee}}, \bibinfo {author} {\bibfnamefont {J.}~\bibnamefont {Lashley}},
  \bibinfo {author} {\bibfnamefont {J.}~\bibnamefont {Park}}, \bibinfo {author}
  {\bibfnamefont {H.}~\bibnamefont {{\AA}gren}}, \bibinfo {author}
  {\bibfnamefont {G.}~\bibnamefont {Aeppli}}, \ and\ \bibinfo {author}
  {\bibfnamefont {A.~V.}\ \bibnamefont {Balatsky}},\ }\href {\doibase
  10.1103/PhysRevX.8.011010} {\bibfield  {journal} {\bibinfo  {journal} {Phys.
  Rev. X}\ }\textbf {\bibinfo {volume} {8}},\ \bibinfo {pages} {011010}
  (\bibinfo {year} {2018})}\BibitemShut {NoStop}%
\bibitem [{\citenamefont {Li}\ \emph {et~al.}(2017)\citenamefont {Li},
  \citenamefont {Li}, \citenamefont {Hu}, \citenamefont {Li},\ and\
  \citenamefont {Fang}}]{li2017dirac}%
  \BibitemOpen
  \bibfield  {author} {\bibinfo {author} {\bibfnamefont {K.}~\bibnamefont
  {Li}}, \bibinfo {author} {\bibfnamefont {C.}~\bibnamefont {Li}}, \bibinfo
  {author} {\bibfnamefont {J.}~\bibnamefont {Hu}}, \bibinfo {author}
  {\bibfnamefont {Y.}~\bibnamefont {Li}}, \ and\ \bibinfo {author}
  {\bibfnamefont {C.}~\bibnamefont {Fang}},\ }\href {\doibase
  10.1103/PhysRevLett.119.247202} {\bibfield  {journal} {\bibinfo  {journal}
  {Phys. Rev. Lett.}\ }\textbf {\bibinfo {volume} {119}},\ \bibinfo {pages}
  {247202} (\bibinfo {year} {2017})}\BibitemShut {NoStop}%
\bibitem [{\citenamefont {McClarty}(2022)}]{mcclarty2021topological}%
  \BibitemOpen
  \bibfield  {author} {\bibinfo {author} {\bibfnamefont {P.}~\bibnamefont
  {McClarty}},\ }\href {\doibase 10.1146/annurev-conmatphys-031620-104715}
  {\bibfield  {journal} {\bibinfo  {journal} {Annu. Rev. Condens. Matter
  Phys.}\ }\textbf {\bibinfo {volume} {13}},\ \bibinfo {pages} {171} (\bibinfo
  {year} {2022})}\BibitemShut {NoStop}%
\bibitem [{\citenamefont {Mook}\ \emph {et~al.}(2021)\citenamefont {Mook},
  \citenamefont {Plekhanov}, \citenamefont {Klinovaja},\ and\ \citenamefont
  {Loss}}]{mook2021interaction}%
  \BibitemOpen
  \bibfield  {author} {\bibinfo {author} {\bibfnamefont {A.}~\bibnamefont
  {Mook}}, \bibinfo {author} {\bibfnamefont {K.}~\bibnamefont {Plekhanov}},
  \bibinfo {author} {\bibfnamefont {J.}~\bibnamefont {Klinovaja}}, \ and\
  \bibinfo {author} {\bibfnamefont {D.}~\bibnamefont {Loss}},\ }\href {\doibase
  10.1103/PhysRevX.11.021061} {\bibfield  {journal} {\bibinfo  {journal} {Phys.
  Rev. X}\ }\textbf {\bibinfo {volume} {11}},\ \bibinfo {pages} {021061}
  (\bibinfo {year} {2021})}\BibitemShut {NoStop}%
\bibitem [{\citenamefont {Yao}\ \emph {et~al.}(2018)\citenamefont {Yao},
  \citenamefont {Li}, \citenamefont {Wang}, \citenamefont {Xue}, \citenamefont
  {Dan}, \citenamefont {Iida}, \citenamefont {Kamazawa}, \citenamefont {Li},
  \citenamefont {Fang},\ and\ \citenamefont {Li}}]{yao2018}%
  \BibitemOpen
  \bibfield  {author} {\bibinfo {author} {\bibfnamefont {W.}~\bibnamefont
  {Yao}}, \bibinfo {author} {\bibfnamefont {C.}~\bibnamefont {Li}}, \bibinfo
  {author} {\bibfnamefont {L.}~\bibnamefont {Wang}}, \bibinfo {author}
  {\bibfnamefont {S.}~\bibnamefont {Xue}}, \bibinfo {author} {\bibfnamefont
  {Y.}~\bibnamefont {Dan}}, \bibinfo {author} {\bibfnamefont {K.}~\bibnamefont
  {Iida}}, \bibinfo {author} {\bibfnamefont {K.}~\bibnamefont {Kamazawa}},
  \bibinfo {author} {\bibfnamefont {K.}~\bibnamefont {Li}}, \bibinfo {author}
  {\bibfnamefont {C.}~\bibnamefont {Fang}}, \ and\ \bibinfo {author}
  {\bibfnamefont {Y.}~\bibnamefont {Li}},\ }\href {\doibase
  10.1038/s41567-018-0213-x} {\bibfield  {journal} {\bibinfo  {journal} {Nat.
  Phys.}\ }\textbf {\bibinfo {volume} {14}},\ \bibinfo {pages} {1011} (\bibinfo
  {year} {2018})}\BibitemShut {NoStop}%
\bibitem [{\citenamefont {Chen}\ \emph {et~al.}(2018)\citenamefont {Chen},
  \citenamefont {Chung}, \citenamefont {Gao}, \citenamefont {Chen},
  \citenamefont {Stone}, \citenamefont {Kolesnikov}, \citenamefont {Huang},\
  and\ \citenamefont {Dai}}]{chen2018topological}%
  \BibitemOpen
  \bibfield  {author} {\bibinfo {author} {\bibfnamefont {L.}~\bibnamefont
  {Chen}}, \bibinfo {author} {\bibfnamefont {J.-H.}\ \bibnamefont {Chung}},
  \bibinfo {author} {\bibfnamefont {B.}~\bibnamefont {Gao}}, \bibinfo {author}
  {\bibfnamefont {T.}~\bibnamefont {Chen}}, \bibinfo {author} {\bibfnamefont
  {M.~B.}\ \bibnamefont {Stone}}, \bibinfo {author} {\bibfnamefont {A.~I.}\
  \bibnamefont {Kolesnikov}}, \bibinfo {author} {\bibfnamefont
  {Q.}~\bibnamefont {Huang}}, \ and\ \bibinfo {author} {\bibfnamefont
  {P.}~\bibnamefont {Dai}},\ }\href {\doibase 10.1103/PhysRevX.8.041028}
  {\bibfield  {journal} {\bibinfo  {journal} {Phys. Rev. X}\ }\textbf {\bibinfo
  {volume} {8}},\ \bibinfo {pages} {041028} (\bibinfo {year}
  {2018})}\BibitemShut {NoStop}%
\bibitem [{\citenamefont {Yuan}\ \emph {et~al.}(2020)\citenamefont {Yuan},
  \citenamefont {Khait}, \citenamefont {Shu}, \citenamefont {Chou},
  \citenamefont {Stone}, \citenamefont {Clancy}, \citenamefont {Paramekanti},\
  and\ \citenamefont {Kim}}]{yuan2020dirac}%
  \BibitemOpen
  \bibfield  {author} {\bibinfo {author} {\bibfnamefont {B.}~\bibnamefont
  {Yuan}}, \bibinfo {author} {\bibfnamefont {I.}~\bibnamefont {Khait}},
  \bibinfo {author} {\bibfnamefont {G.-J.}\ \bibnamefont {Shu}}, \bibinfo
  {author} {\bibfnamefont {F.~C.}\ \bibnamefont {Chou}}, \bibinfo {author}
  {\bibfnamefont {M.~B.}\ \bibnamefont {Stone}}, \bibinfo {author}
  {\bibfnamefont {J.~P.}\ \bibnamefont {Clancy}}, \bibinfo {author}
  {\bibfnamefont {A.}~\bibnamefont {Paramekanti}}, \ and\ \bibinfo {author}
  {\bibfnamefont {Y.-J.}\ \bibnamefont {Kim}},\ }\href {\doibase
  10.1103/PhysRevX.10.011062} {\bibfield  {journal} {\bibinfo  {journal} {Phys.
  Rev. X}\ }\textbf {\bibinfo {volume} {10}},\ \bibinfo {pages} {011062}
  (\bibinfo {year} {2020})}\BibitemShut {NoStop}%
\bibitem [{\citenamefont {Elliot}\ \emph {et~al.}(2021)\citenamefont {Elliot},
  \citenamefont {McClarty}, \citenamefont {Prabhakaran}, \citenamefont
  {Johnson}, \citenamefont {Walker}, \citenamefont {Manuel},\ and\
  \citenamefont {Coldea}}]{elliot2021order}%
  \BibitemOpen
  \bibfield  {author} {\bibinfo {author} {\bibfnamefont {M.}~\bibnamefont
  {Elliot}}, \bibinfo {author} {\bibfnamefont {P.~A.}\ \bibnamefont
  {McClarty}}, \bibinfo {author} {\bibfnamefont {D.}~\bibnamefont
  {Prabhakaran}}, \bibinfo {author} {\bibfnamefont {R.~D.}\ \bibnamefont
  {Johnson}}, \bibinfo {author} {\bibfnamefont {H.~C.}\ \bibnamefont {Walker}},
  \bibinfo {author} {\bibfnamefont {P.}~\bibnamefont {Manuel}}, \ and\ \bibinfo
  {author} {\bibfnamefont {R.}~\bibnamefont {Coldea}},\ }\href {\doibase
  10.1038/s41467-021-23851-0} {\bibfield  {journal} {\bibinfo  {journal} {Nat.
  Commun.}\ }\textbf {\bibinfo {volume} {12}},\ \bibinfo {pages} {3936}
  (\bibinfo {year} {2021})}\BibitemShut {NoStop}%
\bibitem [{\citenamefont {Cai}\ \emph {et~al.}(2021)\citenamefont {Cai},
  \citenamefont {Bao}, \citenamefont {Gu}, \citenamefont {Gao}, \citenamefont
  {Ma}, \citenamefont {Shangguan}, \citenamefont {Si}, \citenamefont {Dong},
  \citenamefont {Wang}, \citenamefont {Wu}, \citenamefont {Lin}, \citenamefont
  {Wang}, \citenamefont {Ran}, \citenamefont {Li}, \citenamefont {Adroja},
  \citenamefont {Xi}, \citenamefont {Yu}, \citenamefont {Li},\ and\
  \citenamefont {Wen}}]{cai2021topological}%
  \BibitemOpen
  \bibfield  {author} {\bibinfo {author} {\bibfnamefont {Z.}~\bibnamefont
  {Cai}}, \bibinfo {author} {\bibfnamefont {S.}~\bibnamefont {Bao}}, \bibinfo
  {author} {\bibfnamefont {Z.-L.}\ \bibnamefont {Gu}}, \bibinfo {author}
  {\bibfnamefont {Y.-P.}\ \bibnamefont {Gao}}, \bibinfo {author} {\bibfnamefont
  {Z.}~\bibnamefont {Ma}}, \bibinfo {author} {\bibfnamefont {Y.}~\bibnamefont
  {Shangguan}}, \bibinfo {author} {\bibfnamefont {W.}~\bibnamefont {Si}},
  \bibinfo {author} {\bibfnamefont {Z.-Y.}\ \bibnamefont {Dong}}, \bibinfo
  {author} {\bibfnamefont {W.}~\bibnamefont {Wang}}, \bibinfo {author}
  {\bibfnamefont {Y.}~\bibnamefont {Wu}}, \bibinfo {author} {\bibfnamefont
  {D.}~\bibnamefont {Lin}}, \bibinfo {author} {\bibfnamefont {J.}~\bibnamefont
  {Wang}}, \bibinfo {author} {\bibfnamefont {K.}~\bibnamefont {Ran}}, \bibinfo
  {author} {\bibfnamefont {S.}~\bibnamefont {Li}}, \bibinfo {author}
  {\bibfnamefont {D.}~\bibnamefont {Adroja}}, \bibinfo {author} {\bibfnamefont
  {X.}~\bibnamefont {Xi}}, \bibinfo {author} {\bibfnamefont {S.-L.}\
  \bibnamefont {Yu}}, \bibinfo {author} {\bibfnamefont {J.-X.}\ \bibnamefont
  {Li}}, \ and\ \bibinfo {author} {\bibfnamefont {J.}~\bibnamefont {Wen}},\
  }\href {\doibase 10.1103/PhysRevB.104.L020402} {\bibfield  {journal}
  {\bibinfo  {journal} {Phys. Rev. B}\ }\textbf {\bibinfo {volume} {104}},\
  \bibinfo {pages} {L020402} (\bibinfo {year} {2021})}\BibitemShut {NoStop}%
\bibitem [{\citenamefont {Scheie}\ \emph {et~al.}(2022)\citenamefont {Scheie},
  \citenamefont {Laurell}, \citenamefont {CcClarty}, \citenamefont {Granroth},
  \citenamefont {Stone}, \citenamefont {Moessner},\ and\ \citenamefont
  {Nagler}}]{scheie22}%
  \BibitemOpen
  \bibfield  {author} {\bibinfo {author} {\bibfnamefont {A.}~\bibnamefont
  {Scheie}}, \bibinfo {author} {\bibfnamefont {P.}~\bibnamefont {Laurell}},
  \bibinfo {author} {\bibfnamefont {P.~A.}\ \bibnamefont {CcClarty}}, \bibinfo
  {author} {\bibfnamefont {G.~E.}\ \bibnamefont {Granroth}}, \bibinfo {author}
  {\bibfnamefont {M.~B.}\ \bibnamefont {Stone}}, \bibinfo {author}
  {\bibfnamefont {R.}~\bibnamefont {Moessner}}, \ and\ \bibinfo {author}
  {\bibfnamefont {S.~E.}\ \bibnamefont {Nagler}},\ }\href {\doibase
  10.1103/PhysRevLett.128.097201} {\bibfield  {journal} {\bibinfo  {journal}
  {Phys. Rev. Lett.}\ }\textbf {\bibinfo {volume} {128}},\ \bibinfo {pages}
  {097201} (\bibinfo {year} {2022})}\BibitemShut {NoStop}%
\bibitem [{\citenamefont {Chumak}\ \emph {et~al.}(2015)\citenamefont {Chumak},
  \citenamefont {Vasyuchka}, \citenamefont {Serga},\ and\ \citenamefont
  {Hillebrands}}]{chumak2015magnon}%
  \BibitemOpen
  \bibfield  {author} {\bibinfo {author} {\bibfnamefont {A.~V.}\ \bibnamefont
  {Chumak}}, \bibinfo {author} {\bibfnamefont {V.~I.}\ \bibnamefont
  {Vasyuchka}}, \bibinfo {author} {\bibfnamefont {A.~A.}\ \bibnamefont
  {Serga}}, \ and\ \bibinfo {author} {\bibfnamefont {B.}~\bibnamefont
  {Hillebrands}},\ }\href {\doibase 10.1038/nphys3347} {\bibfield  {journal}
  {\bibinfo  {journal} {Nat. Phys.}\ }\textbf {\bibinfo {volume} {11}},\
  \bibinfo {pages} {453} (\bibinfo {year} {2015})}\BibitemShut {NoStop}%
\bibitem [{\citenamefont {Wang}\ \emph {et~al.}(2018)\citenamefont {Wang},
  \citenamefont {Zhang},\ and\ \citenamefont {Wang}}]{wang2018topological}%
  \BibitemOpen
  \bibfield  {author} {\bibinfo {author} {\bibfnamefont {X.~S.}\ \bibnamefont
  {Wang}}, \bibinfo {author} {\bibfnamefont {H.~W.}\ \bibnamefont {Zhang}}, \
  and\ \bibinfo {author} {\bibfnamefont {X.~R.}\ \bibnamefont {Wang}},\ }\href
  {\doibase 10.1103/PhysRevApplied.9.024029} {\bibfield  {journal} {\bibinfo
  {journal} {Phys. Rev. Appl.}\ }\textbf {\bibinfo {volume} {9}},\ \bibinfo
  {pages} {024029} (\bibinfo {year} {2018})}\BibitemShut {NoStop}%
\bibitem [{\citenamefont {Pirro}\ \emph {et~al.}(2021)\citenamefont {Pirro},
  \citenamefont {Vasyuchka}, \citenamefont {Serga},\ and\ \citenamefont
  {Hillebrands}}]{pirro2021advances}%
  \BibitemOpen
  \bibfield  {author} {\bibinfo {author} {\bibfnamefont {P.}~\bibnamefont
  {Pirro}}, \bibinfo {author} {\bibfnamefont {V.~I.}\ \bibnamefont
  {Vasyuchka}}, \bibinfo {author} {\bibfnamefont {A.~A.}\ \bibnamefont
  {Serga}}, \ and\ \bibinfo {author} {\bibfnamefont {B.}~\bibnamefont
  {Hillebrands}},\ }\href {\doibase 10.1038/s41578-021-00332-w} {\bibfield
  {journal} {\bibinfo  {journal} {Nat. Rev. Mater.}\ }\textbf {\bibinfo
  {volume} {6}},\ \bibinfo {pages} {1114} (\bibinfo {year} {2021})}\BibitemShut
  {NoStop}%
\bibitem [{\citenamefont {Chen}\ \emph {et~al.}(2021)\citenamefont {Chen},
  \citenamefont {Stone}, \citenamefont {Kolesnikov}, \citenamefont {Winn},
  \citenamefont {Shon}, \citenamefont {Dai},\ and\ \citenamefont
  {Chung}}]{chen2021massless}%
  \BibitemOpen
  \bibfield  {author} {\bibinfo {author} {\bibfnamefont {L.}~\bibnamefont
  {Chen}}, \bibinfo {author} {\bibfnamefont {M.~B.}\ \bibnamefont {Stone}},
  \bibinfo {author} {\bibfnamefont {A.~I.}\ \bibnamefont {Kolesnikov}},
  \bibinfo {author} {\bibfnamefont {B.}~\bibnamefont {Winn}}, \bibinfo {author}
  {\bibfnamefont {W.}~\bibnamefont {Shon}}, \bibinfo {author} {\bibfnamefont
  {P.}~\bibnamefont {Dai}}, \ and\ \bibinfo {author} {\bibfnamefont {J.-H.}\
  \bibnamefont {Chung}},\ }\href {\doibase 10.1088/2053-1583/ac2e7a} {\bibfield
   {journal} {\bibinfo  {journal} {2D Mater.}\ }\textbf {\bibinfo {volume}
  {9}},\ \bibinfo {pages} {015006} (\bibinfo {year} {2021})}\BibitemShut
  {NoStop}%
\bibitem [{\citenamefont {Morosin}\ and\ \citenamefont
  {Narath}(1964)}]{morosin64}%
  \BibitemOpen
  \bibfield  {author} {\bibinfo {author} {\bibfnamefont {B.}~\bibnamefont
  {Morosin}}\ and\ \bibinfo {author} {\bibfnamefont {A.}~\bibnamefont
  {Narath}},\ }\href {\doibase 10.1063/1.1725428} {\bibfield  {journal}
  {\bibinfo  {journal} {J. Chem. Phys.}\ }\textbf {\bibinfo {volume} {40}},\
  \bibinfo {pages} {1958} (\bibinfo {year} {1964})}\BibitemShut {NoStop}%
\bibitem [{\citenamefont {Samuelsen}\ \emph {et~al.}(1971)\citenamefont
  {Samuelsen}, \citenamefont {Silberglitt}, \citenamefont {Shirane},\ and\
  \citenamefont {Remeika}}]{samuelsen1971spin}%
  \BibitemOpen
  \bibfield  {author} {\bibinfo {author} {\bibfnamefont {E.~J.}\ \bibnamefont
  {Samuelsen}}, \bibinfo {author} {\bibfnamefont {R.}~\bibnamefont
  {Silberglitt}}, \bibinfo {author} {\bibfnamefont {G.}~\bibnamefont
  {Shirane}}, \ and\ \bibinfo {author} {\bibfnamefont {J.~P.}\ \bibnamefont
  {Remeika}},\ }\href {\doibase 10.1103/PhysRevB.3.157} {\bibfield  {journal}
  {\bibinfo  {journal} {Phys. Rev. B}\ }\textbf {\bibinfo {volume} {3}},\
  \bibinfo {pages} {157} (\bibinfo {year} {1971})}\BibitemShut {NoStop}%
\bibitem [{\citenamefont {McGuire}\ \emph {et~al.}(2015)\citenamefont
  {McGuire}, \citenamefont {Dixit}, \citenamefont {Cooper},\ and\ \citenamefont
  {Sales}}]{mcguire15}%
  \BibitemOpen
  \bibfield  {author} {\bibinfo {author} {\bibfnamefont {M.~A.}\ \bibnamefont
  {McGuire}}, \bibinfo {author} {\bibfnamefont {H.}~\bibnamefont {Dixit}},
  \bibinfo {author} {\bibfnamefont {V.~R.}\ \bibnamefont {Cooper}}, \ and\
  \bibinfo {author} {\bibfnamefont {B.~A.}\ \bibnamefont {Sales}},\ }\href
  {\doibase 10.1021/cm504242t} {\bibfield  {journal} {\bibinfo  {journal}
  {Chem. Mater.}\ }\textbf {\bibinfo {volume} {27}},\ \bibinfo {pages} {612}
  (\bibinfo {year} {2015})}\BibitemShut {NoStop}%
\bibitem [{\citenamefont {Wang}\ \emph {et~al.}(2011)\citenamefont {Wang},
  \citenamefont {Eyert},\ and\ \citenamefont
  {Schwingenschl{\"o}gl}}]{wang2011electronic}%
  \BibitemOpen
  \bibfield  {author} {\bibinfo {author} {\bibfnamefont {H.}~\bibnamefont
  {Wang}}, \bibinfo {author} {\bibfnamefont {V.}~\bibnamefont {Eyert}}, \ and\
  \bibinfo {author} {\bibfnamefont {U.}~\bibnamefont {Schwingenschl{\"o}gl}},\
  }\href {\doibase 10.1088/0953-8984/23/11/116003} {\bibfield  {journal}
  {\bibinfo  {journal} {J. Phys.: Condens. Matter}\ }\textbf {\bibinfo {volume}
  {23}},\ \bibinfo {pages} {116003} (\bibinfo {year} {2011})}\BibitemShut
  {NoStop}%
\bibitem [{\citenamefont {Do}\ \emph {et~al.}(2022)\citenamefont {Do},
  \citenamefont {Paddison}, \citenamefont {Sala}, \citenamefont {Williams},
  \citenamefont {Kaneko}, \citenamefont {Kuwahara}, \citenamefont {May},
  \citenamefont {Yan}, \citenamefont {McGuire}, \citenamefont {Stone},
  \citenamefont {Lumsden},\ and\ \citenamefont {Christianson}}]{do2022}%
  \BibitemOpen
  \bibfield  {author} {\bibinfo {author} {\bibfnamefont {S.-H.}\ \bibnamefont
  {Do}}, \bibinfo {author} {\bibfnamefont {J.~A.~M.}\ \bibnamefont {Paddison}},
  \bibinfo {author} {\bibfnamefont {G.}~\bibnamefont {Sala}}, \bibinfo {author}
  {\bibfnamefont {T.~J.}\ \bibnamefont {Williams}}, \bibinfo {author}
  {\bibfnamefont {K.}~\bibnamefont {Kaneko}}, \bibinfo {author} {\bibfnamefont
  {K.}~\bibnamefont {Kuwahara}}, \bibinfo {author} {\bibfnamefont {A.~F.}\
  \bibnamefont {May}}, \bibinfo {author} {\bibfnamefont {J.}~\bibnamefont
  {Yan}}, \bibinfo {author} {\bibfnamefont {M.~A.}\ \bibnamefont {McGuire}},
  \bibinfo {author} {\bibfnamefont {M.~D.}\ \bibnamefont {Stone}}, \bibinfo
  {author} {\bibfnamefont {M.~D.}\ \bibnamefont {Lumsden}}, \ and\ \bibinfo
  {author} {\bibfnamefont {A.~D.}\ \bibnamefont {Christianson}},\ }\href
  {https://arxiv.org/abs/2204.03720} {\bibfield  {journal} {\bibinfo  {journal}
  {arXiv:2204.03720}\ } (\bibinfo {year} {2022})}\BibitemShut {NoStop}%
\bibitem [{\citenamefont {Yelon}\ and\ \citenamefont
  {Silberglitt}(1971)}]{yelon1971renormalization}%
  \BibitemOpen
  \bibfield  {author} {\bibinfo {author} {\bibfnamefont {W.~B.}\ \bibnamefont
  {Yelon}}\ and\ \bibinfo {author} {\bibfnamefont {R.}~\bibnamefont
  {Silberglitt}},\ }\href {\doibase 10.1103/PhysRevB.4.2280} {\bibfield
  {journal} {\bibinfo  {journal} {Phys. Rev. B}\ }\textbf {\bibinfo {volume}
  {4}},\ \bibinfo {pages} {2280} (\bibinfo {year} {1971})}\BibitemShut
  {NoStop}%
\bibitem [{SI()}]{SI}%
  \BibitemOpen
  \href@noop {} {}\bibinfo {howpublished} {See the Supplemental Materials at
  http://www.xxx.yyy, which contains Refs.~\cite{xiao2022,dyson56a,dyson56b},
  for a full exposition of our data reduction and fitting, of the bin-width
  error that can appear as a gap at the Dirac point, of the intensity winding
  property at this point, of the comparison with literature results, and of the
  adapted SWT calculations we perform to obtain the first-order magnon
  self-energy in the $J_1$-$J_2$-$J_3$ model.}\BibitemShut {Stop}%
\bibitem [{pan()}]{panther}%
  \BibitemOpen
  \href@noop {} {}\bibinfo {howpublished}
  {https://www.ill.eu/users/instruments/instrument-list/ panther}\BibitemShut
  {NoStop}%
\bibitem [{\citenamefont {Nikitin}\ \emph {et~al.}(2021)\citenamefont
  {Nikitin}, \citenamefont {F{\aa}k}, \citenamefont {Kr\"{a}mer},\ and\
  \citenamefont {R\"{u}egg}}]{ILLdoi}%
  \BibitemOpen
  \bibfield  {author} {\bibinfo {author} {\bibfnamefont {S.}~\bibnamefont
  {Nikitin}}, \bibinfo {author} {\bibfnamefont {B.}~\bibnamefont {F{\aa}k}},
  \bibinfo {author} {\bibfnamefont {K.~W.}\ \bibnamefont {Kr\"{a}mer}}, \ and\
  \bibinfo {author} {\bibfnamefont {C.}~\bibnamefont {R\"{u}egg}},\ }\href
  {\doibase doi:10.5291/ILL-DATA.DIR-236} {\  (\bibinfo {year} {2021}),\
  doi:10.5291/ILL-DATA.DIR-236}\BibitemShut {NoStop}%
\bibitem [{\citenamefont {Stuhr}\ \emph {et~al.}(2017)\citenamefont {Stuhr},
  \citenamefont {Roessli}, \citenamefont {Gvasaliya}, \citenamefont
  {R{\o}nnow}, \citenamefont {Filges}, \citenamefont {Graf}, \citenamefont
  {Bollhalder}, \citenamefont {Hohl}, \citenamefont {B{\"u}rge}, \citenamefont
  {Schild}, \citenamefont {Holitzner}, \citenamefont {Kaegi}, \citenamefont
  {Keller},\ and\ \citenamefont {M\"uhlebach}}]{stuhr2017}%
  \BibitemOpen
  \bibfield  {author} {\bibinfo {author} {\bibfnamefont {U.}~\bibnamefont
  {Stuhr}}, \bibinfo {author} {\bibfnamefont {B.}~\bibnamefont {Roessli}},
  \bibinfo {author} {\bibfnamefont {S.}~\bibnamefont {Gvasaliya}}, \bibinfo
  {author} {\bibfnamefont {H.~M.}\ \bibnamefont {R{\o}nnow}}, \bibinfo {author}
  {\bibfnamefont {U.}~\bibnamefont {Filges}}, \bibinfo {author} {\bibfnamefont
  {D.}~\bibnamefont {Graf}}, \bibinfo {author} {\bibfnamefont {A.}~\bibnamefont
  {Bollhalder}}, \bibinfo {author} {\bibfnamefont {D.}~\bibnamefont {Hohl}},
  \bibinfo {author} {\bibfnamefont {R.}~\bibnamefont {B{\"u}rge}}, \bibinfo
  {author} {\bibfnamefont {M.}~\bibnamefont {Schild}}, \bibinfo {author}
  {\bibfnamefont {L.}~\bibnamefont {Holitzner}}, \bibinfo {author}
  {\bibfnamefont {C.}~\bibnamefont {Kaegi}}, \bibinfo {author} {\bibfnamefont
  {P.}~\bibnamefont {Keller}}, \ and\ \bibinfo {author} {\bibfnamefont
  {T.}~\bibnamefont {M\"uhlebach}},\ }\href {\doibase
  10.1016/j.nima.2017.02.003} {\bibfield  {journal} {\bibinfo  {journal} {Nucl.
  Instrum. Meth. A}\ }\textbf {\bibinfo {volume} {853}},\ \bibinfo {pages} {16}
  (\bibinfo {year} {2017})}\BibitemShut {NoStop}%
\bibitem [{\citenamefont {Arnold}\ \emph {et~al.}(2014)\citenamefont {Arnold},
  \citenamefont {Bilheux}, \citenamefont {Borreguero}, \citenamefont {Buts},
  \citenamefont {Campbell}, \citenamefont {Chapon}, \citenamefont {Doucet},
  \citenamefont {Draper}, \citenamefont {Leal}, \citenamefont {Gigg},
  \citenamefont {Lynch}, \citenamefont {Markvardsen}, \citenamefont
  {Mikkelson}, \citenamefont {Mikkelson}, \citenamefont {Miller}, \citenamefont
  {Palmen}, \citenamefont {Parker}, \citenamefont {Passos}, \citenamefont
  {Perring}, \citenamefont {Peterson}, \citenamefont {Ren}, \citenamefont
  {Reuter}, \citenamefont {Savici}, \citenamefont {Taylor}, \citenamefont
  {Taylor}, \citenamefont {Tolchenov}, \citenamefont {Zhou},\ and\
  \citenamefont {Zikovsky}}]{Mantid}%
  \BibitemOpen
  \bibfield  {author} {\bibinfo {author} {\bibfnamefont {O.}~\bibnamefont
  {Arnold}}, \bibinfo {author} {\bibfnamefont {J.~C.}\ \bibnamefont {Bilheux}},
  \bibinfo {author} {\bibfnamefont {J.~M.}\ \bibnamefont {Borreguero}},
  \bibinfo {author} {\bibfnamefont {A.}~\bibnamefont {Buts}}, \bibinfo {author}
  {\bibfnamefont {S.~I.}\ \bibnamefont {Campbell}}, \bibinfo {author}
  {\bibfnamefont {L.}~\bibnamefont {Chapon}}, \bibinfo {author} {\bibfnamefont
  {M.}~\bibnamefont {Doucet}}, \bibinfo {author} {\bibfnamefont
  {N.}~\bibnamefont {Draper}}, \bibinfo {author} {\bibfnamefont {R.~F.}\
  \bibnamefont {Leal}}, \bibinfo {author} {\bibfnamefont {M.~A.}\ \bibnamefont
  {Gigg}}, \bibinfo {author} {\bibfnamefont {V.~E.}\ \bibnamefont {Lynch}},
  \bibinfo {author} {\bibfnamefont {A.}~\bibnamefont {Markvardsen}}, \bibinfo
  {author} {\bibfnamefont {D.~J.}\ \bibnamefont {Mikkelson}}, \bibinfo {author}
  {\bibfnamefont {R.~L.}\ \bibnamefont {Mikkelson}}, \bibinfo {author}
  {\bibfnamefont {R.}~\bibnamefont {Miller}}, \bibinfo {author} {\bibfnamefont
  {K.}~\bibnamefont {Palmen}}, \bibinfo {author} {\bibfnamefont
  {P.}~\bibnamefont {Parker}}, \bibinfo {author} {\bibfnamefont
  {G.}~\bibnamefont {Passos}}, \bibinfo {author} {\bibfnamefont {T.~G.}\
  \bibnamefont {Perring}}, \bibinfo {author} {\bibfnamefont {P.~F.}\
  \bibnamefont {Peterson}}, \bibinfo {author} {\bibfnamefont {S.}~\bibnamefont
  {Ren}}, \bibinfo {author} {\bibfnamefont {M.~A.}\ \bibnamefont {Reuter}},
  \bibinfo {author} {\bibfnamefont {A.~T.}\ \bibnamefont {Savici}}, \bibinfo
  {author} {\bibfnamefont {J.~W.}\ \bibnamefont {Taylor}}, \bibinfo {author}
  {\bibfnamefont {R.~J.}\ \bibnamefont {Taylor}}, \bibinfo {author}
  {\bibfnamefont {R.}~\bibnamefont {Tolchenov}}, \bibinfo {author}
  {\bibfnamefont {W.}~\bibnamefont {Zhou}}, \ and\ \bibinfo {author}
  {\bibfnamefont {J.}~\bibnamefont {Zikovsky}},\ }\href {\doibase
  10.1016/j.nima.2014.07.029} {\bibfield  {journal} {\bibinfo  {journal} {Nucl.
  Instrum. Meth. A}\ }\textbf {\bibinfo {volume} {764}},\ \bibinfo {pages}
  {156} (\bibinfo {year} {2014})}\BibitemShut {NoStop}%
\bibitem [{\citenamefont {Ewings}\ \emph {et~al.}(2016)\citenamefont {Ewings},
  \citenamefont {Buts}, \citenamefont {Le}, \citenamefont {van Duijn},
  \citenamefont {Bustinduy},\ and\ \citenamefont {Perring}}]{Horace}%
  \BibitemOpen
  \bibfield  {author} {\bibinfo {author} {\bibfnamefont {R.~A.}\ \bibnamefont
  {Ewings}}, \bibinfo {author} {\bibfnamefont {A.}~\bibnamefont {Buts}},
  \bibinfo {author} {\bibfnamefont {M.~D.}\ \bibnamefont {Le}}, \bibinfo
  {author} {\bibfnamefont {J.}~\bibnamefont {van Duijn}}, \bibinfo {author}
  {\bibfnamefont {I.}~\bibnamefont {Bustinduy}}, \ and\ \bibinfo {author}
  {\bibfnamefont {T.~G.}\ \bibnamefont {Perring}},\ }\href {\doibase
  10.1016/j.nima.2016.07.036} {\bibfield  {journal} {\bibinfo  {journal} {Nucl.
  Instrum. Meth. A}\ }\textbf {\bibinfo {volume} {834}},\ \bibinfo {pages}
  {3132} (\bibinfo {year} {2016})}\BibitemShut {NoStop}%
\bibitem [{\citenamefont {Toth}\ and\ \citenamefont {Lake}(2015)}]{toth15}%
  \BibitemOpen
  \bibfield  {author} {\bibinfo {author} {\bibfnamefont {S.}~\bibnamefont
  {Toth}}\ and\ \bibinfo {author} {\bibfnamefont {B.}~\bibnamefont {Lake}},\
  }\href {\doibase 10.1088/0953-8984/27/16/166002} {\bibfield  {journal}
  {\bibinfo  {journal} {J. Phys.: Condens. Matter}\ }\textbf {\bibinfo {volume}
  {27}},\ \bibinfo {pages} {166002} (\bibinfo {year} {2015})}\BibitemShut
  {NoStop}%
\bibitem [{\citenamefont {Alyoshin}\ \emph {et~al.}(1997)\citenamefont
  {Alyoshin}, \citenamefont {Berezin},\ and\ \citenamefont
  {Tulin}}]{alyoshin1997rf}%
  \BibitemOpen
  \bibfield  {author} {\bibinfo {author} {\bibfnamefont {V.}~\bibnamefont
  {Alyoshin}}, \bibinfo {author} {\bibfnamefont {V.}~\bibnamefont {Berezin}}, \
  and\ \bibinfo {author} {\bibfnamefont {V.}~\bibnamefont {Tulin}},\ }\href
  {\doibase 10.1103/PhysRevB.56.719} {\bibfield  {journal} {\bibinfo  {journal}
  {Phys. Rev. B}\ }\textbf {\bibinfo {volume} {56}},\ \bibinfo {pages} {719}
  (\bibinfo {year} {1997})}\BibitemShut {NoStop}%
\bibitem [{\citenamefont {Shivam}\ \emph {et~al.}(2017)\citenamefont {Shivam},
  \citenamefont {Coldea}, \citenamefont {Moessner},\ and\ \citenamefont
  {McClarty}}]{shivam2017neutron}%
  \BibitemOpen
  \bibfield  {author} {\bibinfo {author} {\bibfnamefont {S.}~\bibnamefont
  {Shivam}}, \bibinfo {author} {\bibfnamefont {R.}~\bibnamefont {Coldea}},
  \bibinfo {author} {\bibfnamefont {R.}~\bibnamefont {Moessner}}, \ and\
  \bibinfo {author} {\bibfnamefont {P.}~\bibnamefont {McClarty}},\ }\href
  {https://arxiv.org/abs/1712.08535} {\bibfield  {journal} {\bibinfo  {journal}
  {arXiv:1712.08535}\ } (\bibinfo {year} {2017})}\BibitemShut {NoStop}%
\bibitem [{\citenamefont {Dillon}(1962)}]{dillon1962ferromagnetic}%
  \BibitemOpen
  \bibfield  {author} {\bibinfo {author} {\bibfnamefont {J.}~\bibnamefont
  {Dillon}},\ }in\ \href
  {https://link.springer.com/chapter/10.1007/978-1-4899-6391-8_64} {\emph
  {\bibinfo {booktitle} {Proceedings of the Seventh Conference on Magnetism and
  Magnetic Materials}}}\ (\bibinfo {organization} {Springer},\ \bibinfo {year}
  {1962})\ pp.\ \bibinfo {pages} {1191--1192}\BibitemShut {NoStop}%
\bibitem [{\citenamefont {Bloch}(1930)}]{bloch30}%
  \BibitemOpen
  \bibfield  {author} {\bibinfo {author} {\bibfnamefont {F.}~\bibnamefont
  {Bloch}},\ }\href {\doibase 10.1007/BF01339661} {\bibfield  {journal}
  {\bibinfo  {journal} {Z. Phys.}\ }\textbf {\bibinfo {volume} {61}},\ \bibinfo
  {pages} {206} (\bibinfo {year} {1930})}\BibitemShut {NoStop}%
\bibitem [{\citenamefont {Becker}\ and\ \citenamefont
  {Wessel}(2018)}]{becker18}%
  \BibitemOpen
  \bibfield  {author} {\bibinfo {author} {\bibfnamefont {J.}~\bibnamefont
  {Becker}}\ and\ \bibinfo {author} {\bibfnamefont {S.}~\bibnamefont
  {Wessel}},\ }\href {\doibase 10.1103/PhysRevLett.121.077202} {\bibfield
  {journal} {\bibinfo  {journal} {Phys. Rev. Lett.}\ }\textbf {\bibinfo
  {volume} {121}},\ \bibinfo {pages} {077202} (\bibinfo {year}
  {2018})}\BibitemShut {NoStop}%
\bibitem [{\citenamefont {Shao}\ and\ \citenamefont {Sandvik}(2022)}]{shao22}%
  \BibitemOpen
  \bibfield  {author} {\bibinfo {author} {\bibfnamefont {H.}~\bibnamefont
  {Shao}}\ and\ \bibinfo {author} {\bibfnamefont {A.~W.}\ \bibnamefont
  {Sandvik}},\ }\href {https://arxiv.org/abs/2202.09870} {\bibfield  {journal}
  {\bibinfo  {journal} {arXiv:2202.09870}\ } (\bibinfo {year}
  {2022})}\BibitemShut {NoStop}%
\bibitem [{\citenamefont {Zauner-Stauber}\ \emph {et~al.}(2018)\citenamefont
  {Zauner-Stauber}, \citenamefont {Vanderstraeten}, \citenamefont {Haegeman},
  \citenamefont {McCulloch},\ and\ \citenamefont {Verstraete}}]{zauner18}%
  \BibitemOpen
  \bibfield  {author} {\bibinfo {author} {\bibfnamefont {V.}~\bibnamefont
  {Zauner-Stauber}}, \bibinfo {author} {\bibfnamefont {L.}~\bibnamefont
  {Vanderstraeten}}, \bibinfo {author} {\bibfnamefont {J.}~\bibnamefont
  {Haegeman}}, \bibinfo {author} {\bibfnamefont {I.~P.}\ \bibnamefont
  {McCulloch}}, \ and\ \bibinfo {author} {\bibfnamefont {F.}~\bibnamefont
  {Verstraete}},\ }\href {\doibase 10.1103/PhysRevB.97.235155} {\bibfield
  {journal} {\bibinfo  {journal} {Phys. Rev. B}\ }\textbf {\bibinfo {volume}
  {97}},\ \bibinfo {pages} {235155} (\bibinfo {year} {2018})}\BibitemShut
  {NoStop}%
\bibitem [{\citenamefont {Ponsioen}\ \emph {et~al.}(2022)\citenamefont
  {Ponsioen}, \citenamefont {Assaad},\ and\ \citenamefont
  {Corboz}}]{ponsioen22}%
  \BibitemOpen
  \bibfield  {author} {\bibinfo {author} {\bibfnamefont {B.}~\bibnamefont
  {Ponsioen}}, \bibinfo {author} {\bibfnamefont {F.~F.}\ \bibnamefont
  {Assaad}}, \ and\ \bibinfo {author} {\bibfnamefont {P.}~\bibnamefont
  {Corboz}},\ }\href {\doibase 10.21468/SciPostPhys.12.1.006} {\bibfield
  {journal} {\bibinfo  {journal} {SciPost Phys.}\ }\textbf {\bibinfo {volume}
  {12}},\ \bibinfo {pages} {006} (\bibinfo {year} {2022})}\BibitemShut
  {NoStop}%
\bibitem [{\citenamefont {Xiao}\ \emph {et~al.}(2022)\citenamefont {Xiao},
  \citenamefont {Ma}, \citenamefont {Bryan}, \citenamefont {Fu}, \citenamefont
  {Mann}, \citenamefont {Winn}, \citenamefont {Abernathy}, \citenamefont
  {Hermann}, \citenamefont {Khanolkar}, \citenamefont {Dennett}, \citenamefont
  {Hurley}, \citenamefont {Manley},\ and\ \citenamefont
  {Marianetti}}]{xiao2022}%
  \BibitemOpen
  \bibfield  {author} {\bibinfo {author} {\bibfnamefont {E.}~\bibnamefont
  {Xiao}}, \bibinfo {author} {\bibfnamefont {H.}~\bibnamefont {Ma}}, \bibinfo
  {author} {\bibfnamefont {M.~S.}\ \bibnamefont {Bryan}}, \bibinfo {author}
  {\bibfnamefont {L.}~\bibnamefont {Fu}}, \bibinfo {author} {\bibfnamefont
  {J.~M.}\ \bibnamefont {Mann}}, \bibinfo {author} {\bibfnamefont
  {B.}~\bibnamefont {Winn}}, \bibinfo {author} {\bibfnamefont {D.~L.}\
  \bibnamefont {Abernathy}}, \bibinfo {author} {\bibfnamefont {R.~P.}\
  \bibnamefont {Hermann}}, \bibinfo {author} {\bibfnamefont {A.~R.}\
  \bibnamefont {Khanolkar}}, \bibinfo {author} {\bibfnamefont {C.~A.}\
  \bibnamefont {Dennett}}, \bibinfo {author} {\bibfnamefont {D.~H.}\
  \bibnamefont {Hurley}}, \bibinfo {author} {\bibfnamefont {M.~E.}\
  \bibnamefont {Manley}}, \ and\ \bibinfo {author} {\bibfnamefont {C.~A.}\
  \bibnamefont {Marianetti}},\ }\href {https://arxiv.org/abs/2202.11401}
  {\bibfield  {journal} {\bibinfo  {journal} {arXiv:2202.11041}\ } (\bibinfo
  {year} {2022})}\BibitemShut {NoStop}%
\bibitem [{\citenamefont {Dyson}(1956{\natexlab{a}})}]{dyson56a}%
  \BibitemOpen
  \bibfield  {author} {\bibinfo {author} {\bibfnamefont {F.~J.}\ \bibnamefont
  {Dyson}},\ }\href {\doibase 10.1103/PhysRev.102.1217} {\bibfield  {journal}
  {\bibinfo  {journal} {Phys. Rev.}\ }\textbf {\bibinfo {volume} {102}},\
  \bibinfo {pages} {1217} (\bibinfo {year} {1956}{\natexlab{a}})}\BibitemShut
  {NoStop}%
\bibitem [{\citenamefont {Dyson}(1956{\natexlab{b}})}]{dyson56b}%
  \BibitemOpen
  \bibfield  {author} {\bibinfo {author} {\bibfnamefont {F.~J.}\ \bibnamefont
  {Dyson}},\ }\href {\doibase 10.1103/PhysRev.102.1230} {\bibfield  {journal}
  {\bibinfo  {journal} {Phys. Rev.}\ }\textbf {\bibinfo {volume} {102}},\
  \bibinfo {pages} {1230} (\bibinfo {year} {1956}{\natexlab{b}})}\BibitemShut
  {NoStop}%
\end{thebibliography}%

\clearpage

\onecolumngrid

\setcounter{equation}{0}
\renewcommand{\theequation}{S\arabic{equation}}
\setcounter{figure}{0}
\renewcommand{\thefigure}{S\arabic{figure}}
\setcounter{section}{0}
\renewcommand{\thesection}{S\arabic{section}}
\setcounter{table}{0}
\renewcommand{\thetable}{S\arabic{table}}

\vskip12mm

\centerline{\large {\bf {Supplemental Material for ``Thermal Evolution of 
Dirac Magnons}}} 

\vskip1mm

\centerline{\large {\bf {in the Honeycomb Ferromagnet CrBr$_3$''}}}

\vskip4mm

\centerline{S. E. Nikitin, B. F{\aa}k, K. W. Kr\"amer, T. Fennell, B. Normand, 
A. M. L\"auchli, and Ch. R\"uegg}

\vskip8mm

\twocolumngrid

\section{Sample preparation and characterization}

\subsection{Crystal Growth}

The crystal was prepared from CrBr$_3$ (Cerac, 3N), which first was sublimed 
for purification in a sealed silica ampoule at 700$^\circ$ C under vacuum. 
For crystal growth, the purified material was sealed in a silica ampoule 
under vacuum and heated to 850$^\circ$ C in a vertical tube furnace with a 
small temperature gradient. The crystal grew from the cold region over a 
period of three weeks. Afterwards, the ampoule was cooled to room temperature 
at a rate of 10 K per hour. All handling of the  material was done under dry 
conditions in a glove box or in sealed sample containers.

\subsection{Mosaicity effects}
\label{sec:mos}

\begin{figure}[t]
\includegraphics[width=1\columnwidth]{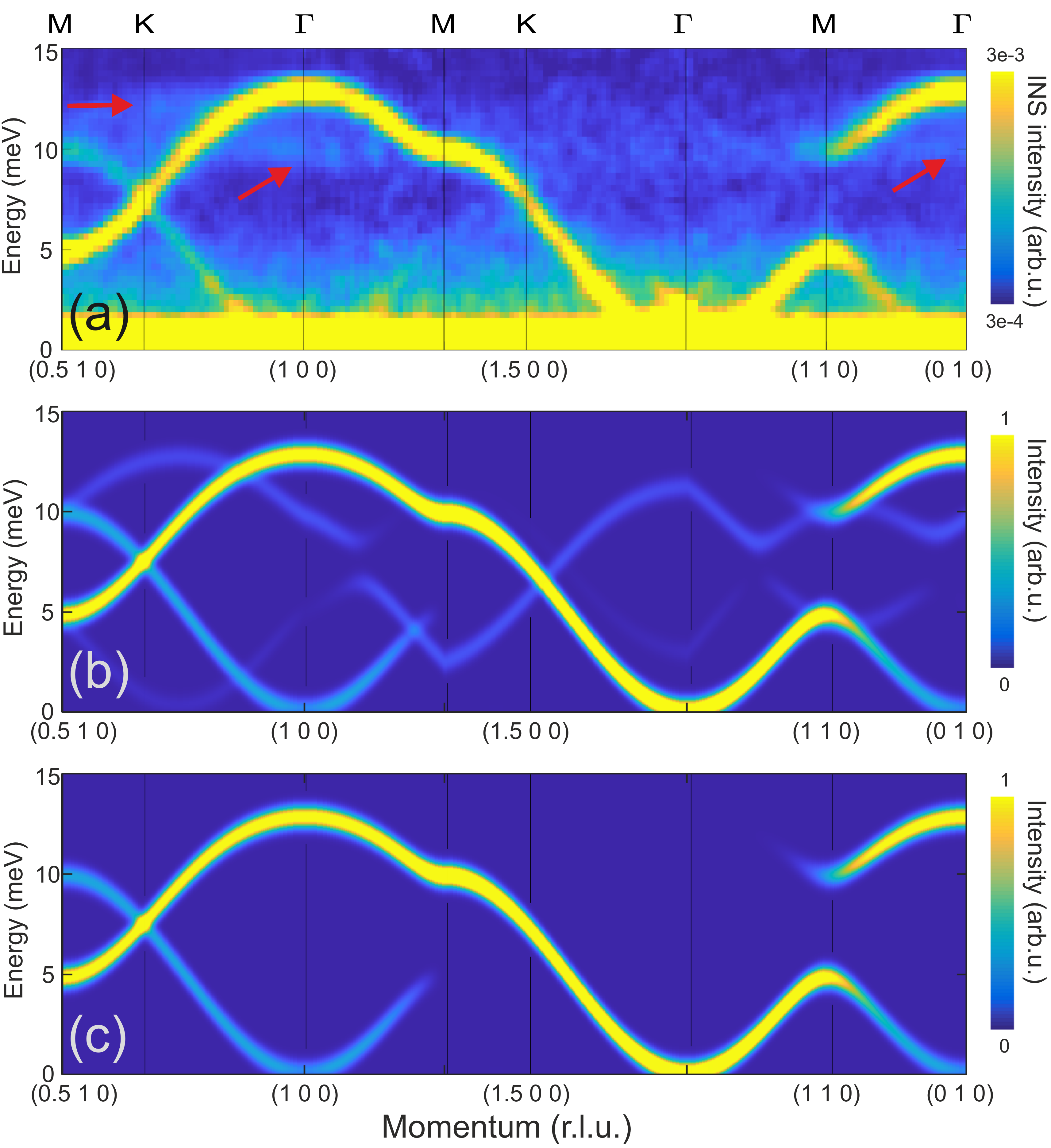}
\caption{(a) Magnetic excitation spectra for the primary high-symmetry 
reciprocal-space directions in \crbr\ taken from the PANTHER $E_{\mathrm{i}}
 = 30$~meV dataset at $T = 1.7$~K. This panel is reproduced from Fig.~3(a) 
of the main text and the red arrows indicate regions contaminated by the 
twin contribution. (b,c) Magnetic excitation spectrum calculated by LSWT 
with (b) and without (c) the twin contribution.}
\label{fig_twin}
\end{figure}

The quality of the sample was characterized by neutron diffraction. We found 
that, in addition to the primary crystallite, it contained a structural twin 
whose in-plane axes were rotated by 30$^{\circ}$ around the $c$ axis. From the 
ratio of the Bragg-peak intensities, we estimated that this twin constituted 
approximately 10\% of the sample mass. To estimate the effects of such a twin 
on the INS spectra shown in Fig.~3(a) of the main text, which we reproduce in 
Fig.~\ref{fig_twin}(a), we modelled the magnetic response of a composite 
system consisting of the twin and the main crystallite for each of the 
high-symmetry paths. As shown in Figs.~\ref{fig_twin}(b) and \ref{fig_twin}(c), 
the twin produces two additional, faint magnon modes, whose traces can also be 
identified in our data [Fig.~\ref{fig_twin}(a)]. Thus we took the twin 
contribution into account when extracting the positions and line widths of the 
measured magnon modes, by fitting it with a separate peak function at low 
temperatures, albeit with the relative positions, widths, and intensities 
of the twin-related peaks all fixed.

\section{INS experiment and data analysis} 
\label{sec:data}

PANTHER is a direct-geometry TOF spectrometer at the ILL. The available 
beam time allowed us to collect data at $T = 1.7$, 20, 30, and 40~K, each 
with two incident neutron energies, $E_{\mathrm{i}} = 15 $ and 30~meV, by 
rotating the crystal through $340^{\circ}$ in steps of 1$^{\circ}$. Spectra 
were collected for 4 minutes at each angle with $E_{\mathrm{i}} = 15$ meV 
and 3.5 minutes at 30~meV. The energy resolution, defined as the full 
width at half-maximum (FWHM) peak height at zero energy transfer were 
respectively 0.58 and 0.79~meV for $E_{\mathrm{i}} = 15$ and $30$~meV.
Reduction and analysis of the TOF data were 
performed using the software \textsc{MANTID} \cite{Mantid} and 
\textsc{HORACE} \cite{Horace}, and a symmetrization procedure was applied 
in order to improve the statistics. The result of this processing was a 
``four-dimensional'' dataset of scattered intensities as a function of 
momentum transfer, $\mathbf{Q}$, and energy transfer, $\omega$, which we 
manipulate for different purposes in the remainder of this section. 

\begin{figure*}[t]
\includegraphics[width=0.96\textwidth]{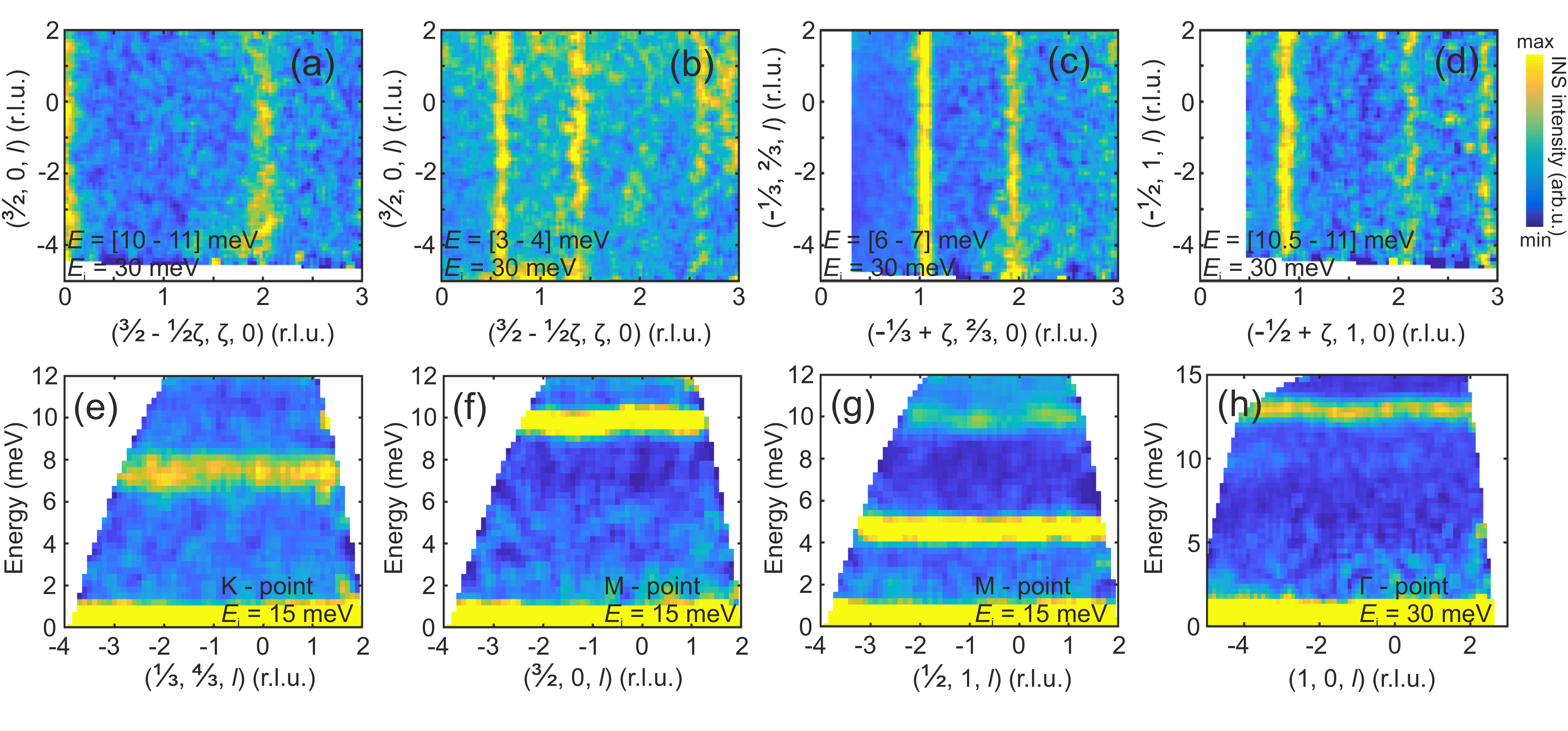}
\caption{(a-d) Selected constant-energy cuts through the INS spectral 
function measured on PANTHER, which show essentially dispersionless 
behavior of the magnetic excitations in the out-of-plane direction, $l$. 
Data were integrated by $\pm 0.03$ r.l.u.~in the orthogonal direction 
and over the energy ranges indicated; the intensity scales were optimized 
separately for each panel. (e-h) INS spectra taken at the high-symmetry 
K (e), M (f,g), and $\Gamma$ (h) points of the 2D BZ also show minimal 
dependence $l$; these intensity data were integrated by $\pm 0.03$ 
r.l.u.~in the orthogonal directions and are shown on a common scale.}
\label{fig_out_of_plane}
\end{figure*}

EIGER is a thermal-neutron TAS at the PSI. We used the fixed-$k_{\mathrm{f}}$ 
mode with $k_{\mathrm{f}} = 2.662$ \AA$^{-1}$, which yields an energy resolution 
(FWHM) of 0.63~meV at zero energy transfer, and collected data at eight 
different temperatures from 1.5 to 40 K. We set horizontal focusing on the 
analyzer and double focusing on the monochromator, also installing a graphite 
filter before the sample to reduce the contamination from second-order 
neutrons. We fitted the constant-$\mathbf{Q}$ cuts shown in Fig.~4 of the main 
text by using Eq.~\eqref{Eq:fit}, given in Sec.~\ref{sec:fit}, in order to 
extract the center and width of the Lorentzian peak.

\subsection{Out-of-plane dispersion}
\label{Sec:out_of_plane}

Although the \crbr\ sample was oriented in the $(h~k~0)$ scattering plane, 
the large vertical coverage of the position-sensitive detectors on PANTHER 
made it possible to collect some data from out-of-plane scattering. Figures 
\ref{fig_out_of_plane}(a-d) show constant-energy cuts through the 30 meV 
dataset taken at four different energies and transverse momentum transfers, 
in which the magnon branches exhibit no measurable dispersion as a function 
of $l$. Figures \ref{fig_out_of_plane}(e-h) show $\mathbf{Q}$-$E$ 
cuts along the $l$ direction taken at selected high-symmetry points, in 
which the magnon branches again appear almost completely flat. A very minor 
modulation at the upper band edge is consistent with fact that the ordering 
temperature, $T_c = 32$ K, implies a weak interplane interaction, $J_\perp$, 
which the instrumental resolution prevents us from fitting accurately. 

\subsection{Magnetic excitations near the Dirac point}

\begin{figure*}[t]
\includegraphics[width=\textwidth]{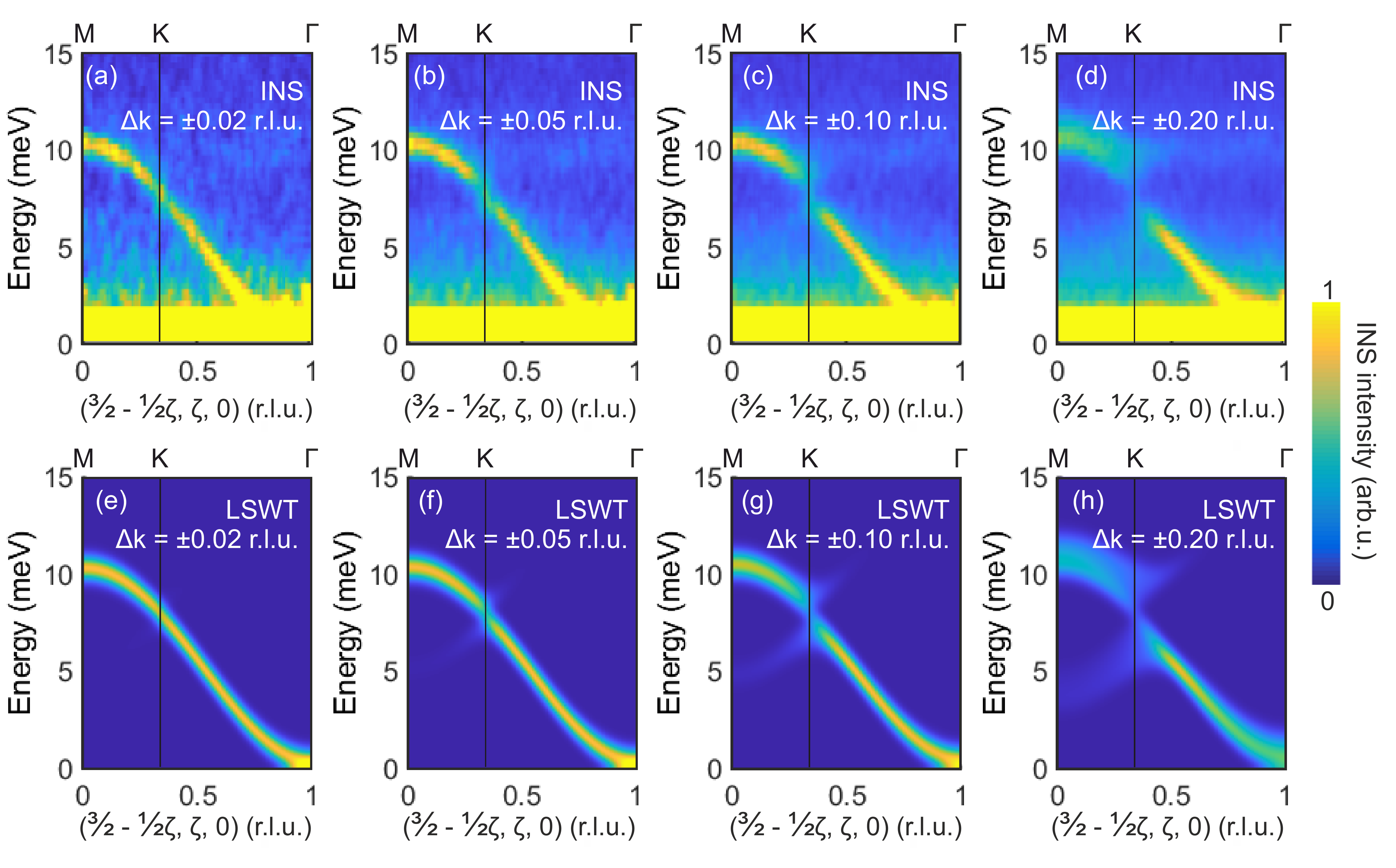}
\caption{INS spectrum of \crbr\ along the $(\frac{3}{2} - \frac12 \xi, \xi, 0)$ 
direction. Panels (a-d) show results obtained by integrating the measured 
intensities over different widths ($\Delta k$) in the $k$ direction and panels 
(e-h) show the analogous results modelled within LSWT. It is clear that the 
spectrum is entirely continuous as $\Delta k$ is taken to zero, with only a 
dip in intensity at the K point arising due to the magnon density of states, 
whereas increasing $\Delta k$ causes the appearance of a spin gap at K. (i) 
Corresponding line widths extracted by using different values of $\Delta k$.}
\label{fig_integration}
\end{figure*}

\crbr\ has for 50 years been regarded as a prototypical honeycomb Heisenberg 
FM \cite{samuelsen1971spin,yelon1971renormalization}, and for this reason 
was used as the test-case material in a recent analysis of the consequences 
of the Dirac cones in its magnon spectrum for its thermal and topological 
properties. Thus the recent work Cai and coauthors \cite{cai2021topological} 
reporting a massive spin gap at the K point, of order 2~meV in a total band 
width of 13 meV, both contradicted previous results and would have profound 
consequences for the physical understanding of \crbr. Here we show that the 
conclusion of Ref.~\cite{cai2021topological} is contradicted by our data and 
we illustrate the most probable reason for this disagreement.

Like us, the authors of Ref.~\cite{cai2021topological} performed TOF INS 
experiments and in Figs.~1 and 3 of their manuscript present the spectra 
measured at their base temperature INS spectra by preparing $\mathbf{Q}$-$E$ 
cuts through the four-dimensional TOF datasets. In this type of analysis, it 
is necessary to choose an integration width in the two orthogonal directions 
in reciprocal space ($\mathbf{Q}$), and for this the authors used $\pm 0.2$ 
r.l.u.~for the in-plane direction and $\pm 5$ out of plane. In this process, 
choosing a large window captures a higher intensity, allowing one to improve 
the statistics, but it can produce spurious features in the prepared cuts if 
the mode does disperse within the selected integration window. To illustrate 
this effect, in Figs.~\ref{fig_integration}(a-d) we show multiple cuts from 
our PANTHER dataset ($E_{\mathrm{i}} = 30$ meV, $T = 1.7$~K) prepared by 
taking different integration widths, $\Delta{}k$, in the orthogonal 
in-plane direction. The spectrum we obtain with $\Delta{}k = \pm 0.2$ 
r.l.u.~[Fig.~\ref{fig_integration}(d)] gives the appearance of a robust 
splitting into separate acoustic and optical branches. However, if we 
decrease $\Delta{}k$ then this apparent gap also decreases, and with our 
data it is clear that there is no splitting for $\Delta{}k \leq 0.05$ 
r.l.u.~[Figs.~\ref{fig_integration}(a-b)].

To further support this observation, we have modelled the INS response of 
\crbr\ by performing LSWT calculations within \textsc{SpinW} \cite{toth15} 
using the spin Hamiltonian and Heisenberg interaction parameters given in 
and below Eq.~(2) of the main text. In Figs.~\ref{fig_integration}(e-h) 
we illustrate the results obtained by integrating the modelled intensity 
over the same ranges of orthogonal $\mathbf{q}$ as we did for the TOF data. 
It is clear that this modelling provides a perfect reproduction of the 
integrated INS data, supporting our conclusion that the splitting reported 
in Ref.~\cite{cai2021topological} is not a property of \crbr, but purely 
an extrinsic consequence of the data treatment. 

The explanation of this effect is straighforward: the integration averages 
multiple intensity pixels in directions perpendicular to the cut. If there 
is little or no orthogonal dispersion across the integration window, the 
procedure is reliable. However, in a 2D honeycomb FM with a suspected Dirac 
cone at the K point, the only dispersionless direction is $l$; in $h$ and 
$k$, all the magnon intensity near the apex of the cone is concentrated in 
a small $\mathbf{Q}$ volume, and one should also note that this intensity 
is weak (the magnon density of states vanishes at a true Dirac point). 
A broad integration window risks missing the apical intensity and generates 
extra intensity at K at finite energies above and below the Dirac point, 
creating the appearance of broadened and separated magnon branches. 

This integration-induced broadening also causes an apparent 
$\mathbf{Q}$-dependence of the magnon line width~\cite{xiao2022}, an effect 
that can become substantial in parts of the BZ with strong dispersion. We 
modelled this broadening with $\textsc{SpinW}$ and found that, for the 
in-plane integration widths we use ($\Delta h = \Delta k = \pm 0.03$ r.l.u.), 
it reaches 20\% for the steepest parts of the magnon dispersion and rises 
rapidly if the integration width is increased. In preparing the results 
presented in Fig.~3 of the main text, we took both the instrumental 
resolution and this integration-induced broadening into account. 

\begin{figure}[b]
\includegraphics[width=1\columnwidth]{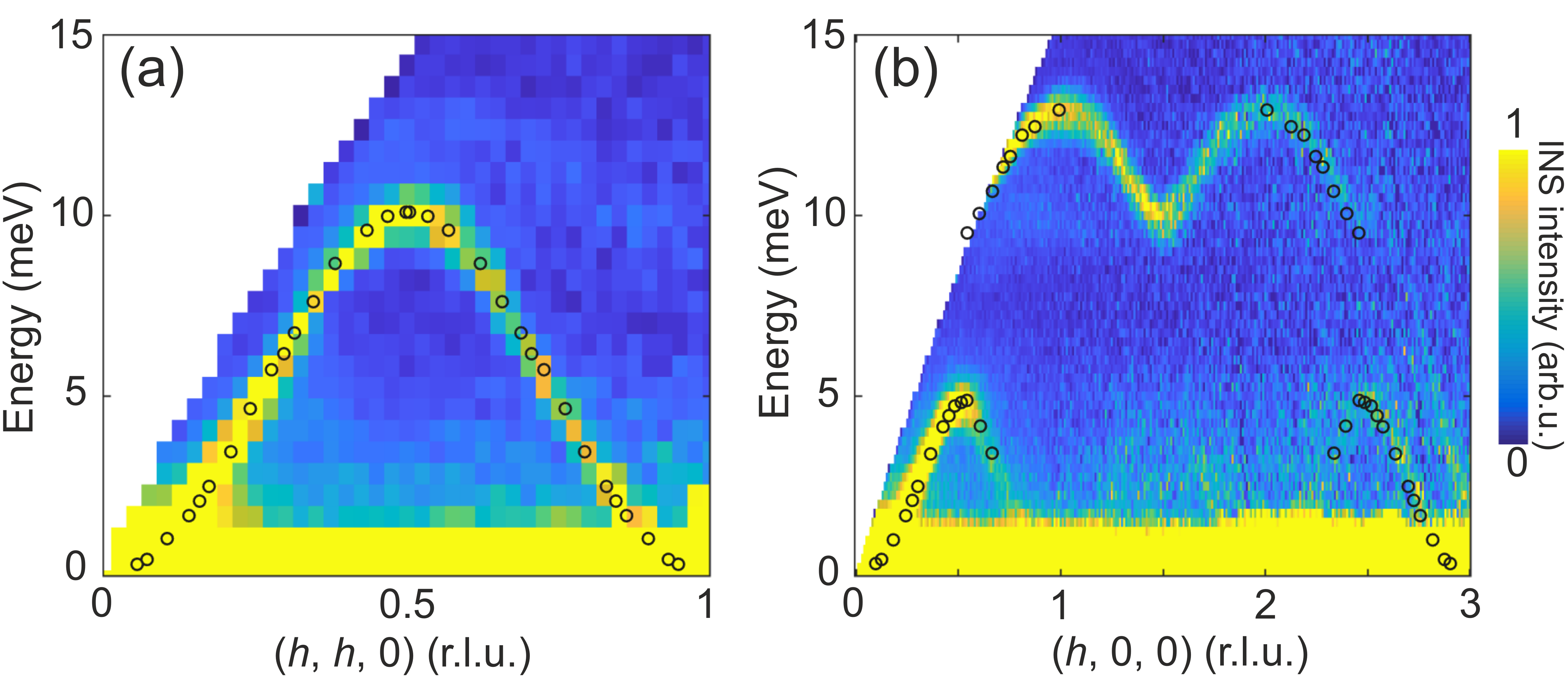}
\caption{INS spectra prepared from the $E_{\mathrm{i}} = 30$~meV PANTHER 
dataset at $T = 1.7$~K for the high-symmetry directions $(h~0~0)$ (a) and 
$(h~h~0)$ (b). Points indicate experimental data taken from Fig.~6 of 
Ref.~\cite{samuelsen1971spin}, which shows results obtained from measurements 
performed at a base temperature of 6 K.}
\label{fig_compare}
\end{figure}

\subsection{Comparison with previous INS studies}

The magnon spectrum of \crbr\ was measured in a series of INS experiments 
in the early 1970s \cite{samuelsen1971spin,yelon1971renormalization}. Because 
a modern measurement of this spectrum \cite{cai2021topological} contradicts 
these old results, but is also contradicted by our data (previous subsection), 
it is worthwhile to compare our results directly with the TAS data from 
Refs.~\cite{samuelsen1971spin, yelon1971renormalization}. In 
Fig.~\ref{fig_compare} we show on top of our PANTHER data the magnon 
center positions deduced by these authors at 6~K for the $(0~0~0) \rightarrow 
(1~1~0)$ and $(0~0~0) \rightarrow (3~0~0)$ paths in the BZ (which correspond 
to the $\Delta$ and $\Sigma$ directions in the notation of 
Ref.~\cite{samuelsen1971spin}). We conclude that the agreement is perfect 
within the resolutions of both measurements. 

\subsection{Intensity winding of Dirac magnons}
 
\begin{figure}[t]
\includegraphics[width=1\columnwidth]{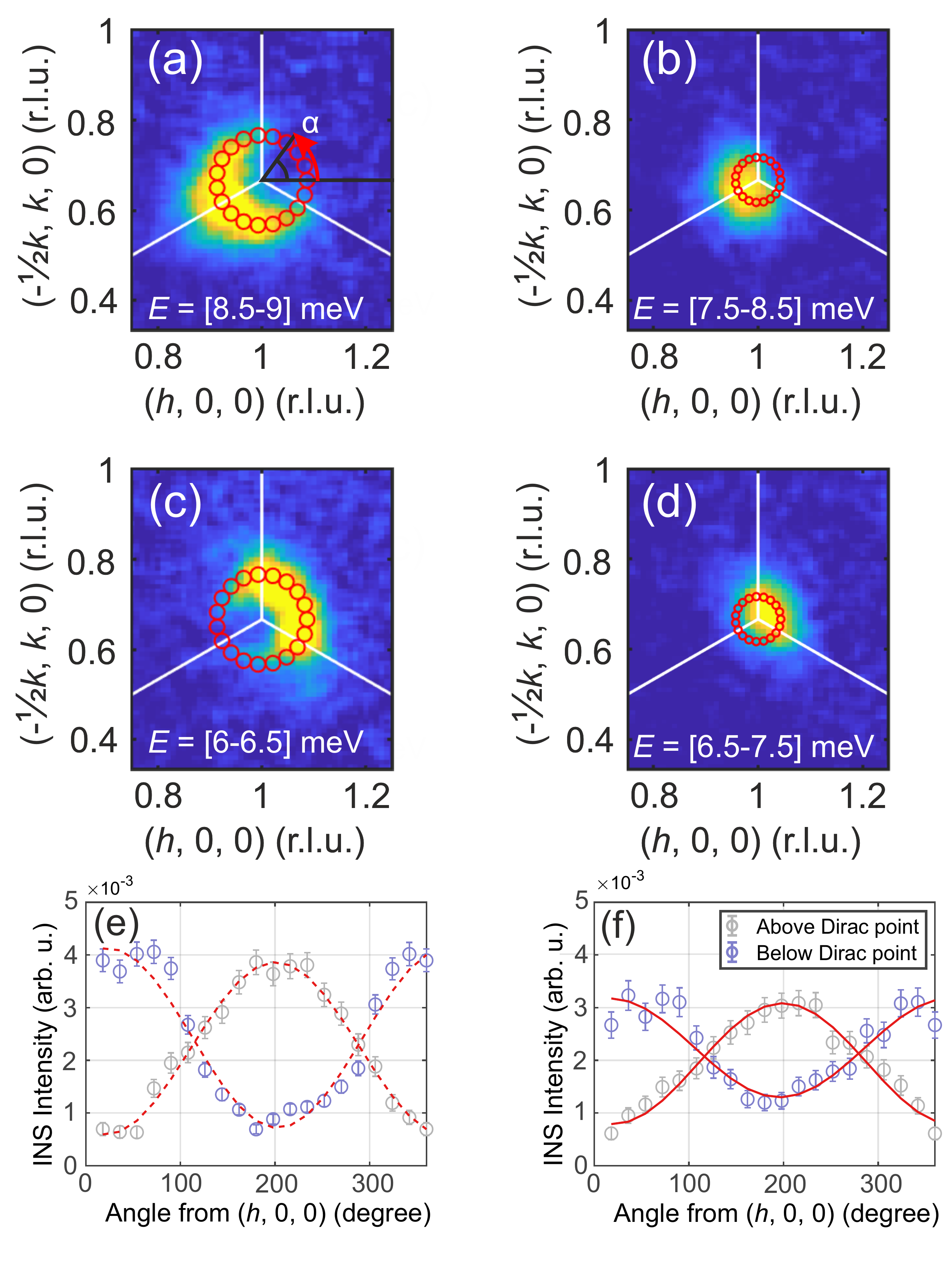}
\caption{(a-d) Constant-energy cuts through the INS spectral function at 
energies above and below the Dirac-point energy of 7.5 meV. Data were 
taken from the $E_{\mathrm{i}} = 30$~meV PANTHER measurements at $T = 
1.7$~K with integration performed over the momentum range $l = \pm 5$ and 
over the energy ranges shown. White lines show boundaries of the BZ and 
the axes in panel (a) show definition of $\alpha$ in Eq.~\eqref{Eq:wind}.
(e) Intensity winding around the K point, 
shown as a function of angle on the path indicated in panels (a) and (c), 
which was chosen at $|\mathbf{q}| = 0.08$ \AA$^{-1}$ away from $\mathbf{q} = 
(2/3~2/3~0)$; this panel is also shown in Fig. 1(d) of the main text. (f) 
Intensity winding around the K point, shown as a function of angle on the 
path indicated in panels (b) and (d), which was chosen at $|\mathbf{q}| = 
0.04$ \AA$^{-1}$ away from $\mathbf{q} = (2/3~2/3~0)$.}
\label{fig_winding}
\end{figure}

A recent theoretical analysis of Dirac and Weyl magnons predicted that the 
dynamical structure factor should exhibit a characteristic type of behavior 
in the vicinity of the special points \cite{shivam2017neutron}. Using the 
honeycomb FM as their first example, these authors showed that the intensity 
should follow
\begin{align}
I = I_0 [1 \pm\ \mathrm{cos}(\alpha - \alpha_0)],
\label{Eq:wind}
\end{align}
on a circle taken in $\mathbf{Q}$ around the K point in the $l = 0$ hexagonal 
plane; here $\alpha$ is the polar angle measured from $\alpha_0$ parallel to 
the $(h,0,0)$ direction [Fig.~\ref{fig_winding}(a)] and the $\pm$ 
sign refers to the magnon bands above and below the Dirac point in energy. 
The origin of this behavior lies in the rotation of the isospin polarization 
of the magnon band, and examples have been observed very recently in 
experiments on CoTiO$_3$, a honeycomb magnet with bond-dependent interactions 
\cite{elliot2021order}, and on elemental Gd, which has a hexagonal close-packed 
structure and RKKY-type magnetic interactions \cite{scheie22}. 

Figure 1(d) of the main text shows a near-ideal cosinusoidal intensity 
winding. These data were obtained from the constant-energy cuts shown in 
Figs.~\ref{fig_winding}(a) and \ref{fig_winding}(c), which were taken 
respectively above and below the K point by integrating over energy windows 
of width 0.5 meV. In both cuts the INS intensity is concentrated around the K 
points and distributed over a semi-circular trajectory, but with peaks on 
the opposite sides of the K points for the upper and lower magnon bands. To 
quantify this effect and to compare it with theory [Eq.~\eqref{Eq:wind}], we 
made azimuthal scans on the trajectory around the K point at $(2/3~2/3~0)$ 
that is indicated in both panels. It is clear from the curves in Fig. 1(d) 
of the main text, which are reproduced in Fig.~\ref{fig_winding}(e), that 
the INS intensities on the bands above and below the Dirac point exhibit the 
anticipated cosine modulation with exactly opposing phases. In 
Figs.~\ref{fig_winding}(b), \ref{fig_winding}(d), and \ref{fig_winding}(f) 
we show that this result is not particularly sensitive to the width of the 
integration window, at least in the regime of linear dispersion, although 
the trajectory should be redefined to match the shape of the Dirac cone. 
This perfect agreement with Eq.~\eqref{Eq:wind} allows us to conclude 
that \crbr\ provides an excellent realization of the isospin winding of 
nodal quasiparticles.

\begin{figure}[t]
\includegraphics[width=1\columnwidth]{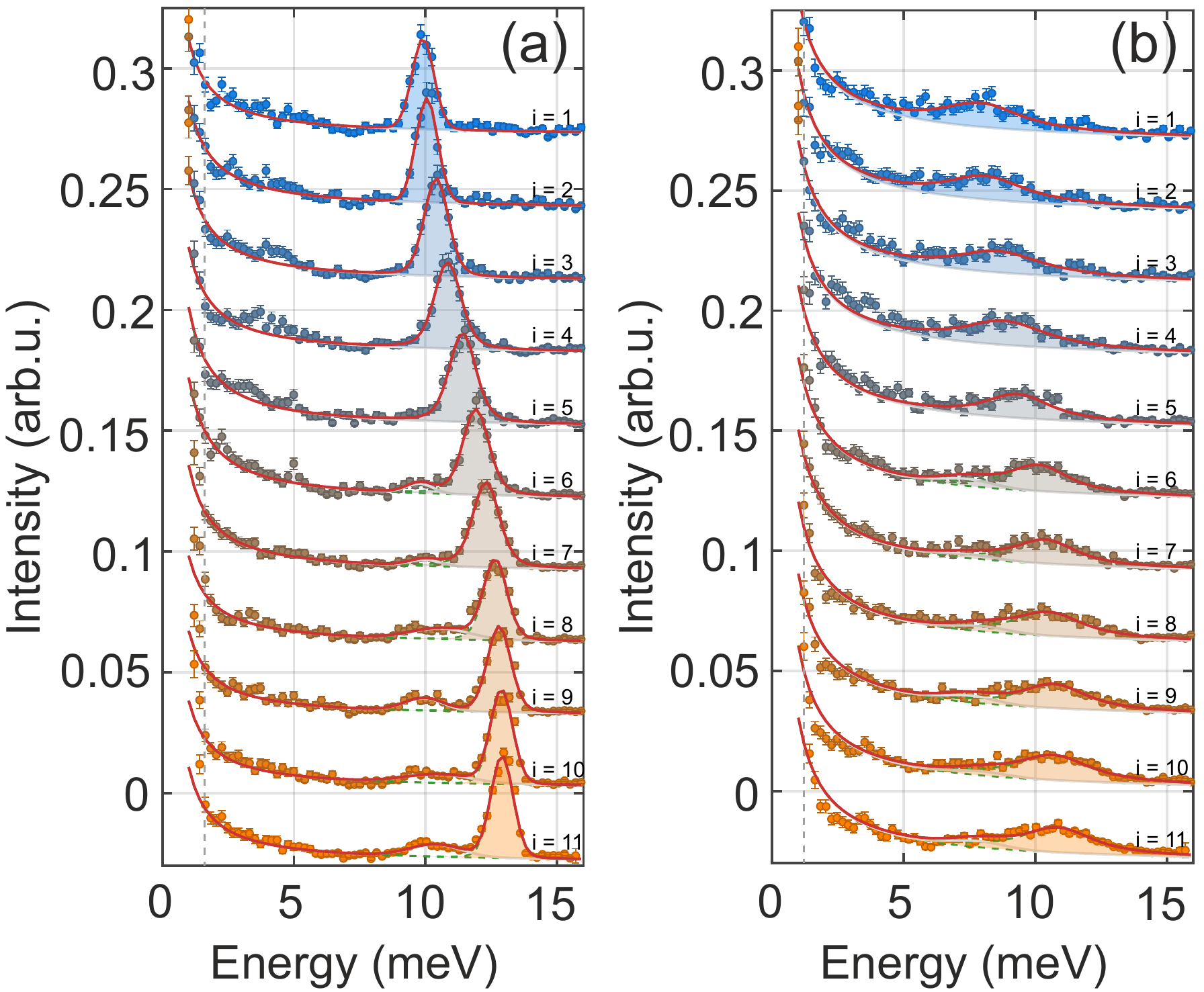}
\caption{Constant-$\mathbf{Q}$ cuts taken from the $E_{\mathrm{i}} = 30$~meV 
PANTHER datasets betewen $\mathbf{Q} = (1.5~0~0)$ ($i = 1$) and $\mathbf{Q} = 
(1~0~0)$ ($i = 11$) in steps of 0.05 r.l.u. Data were integrated over $\pm 
0.03$ r.l.u.~in the orthogonal in-plane directions and by $l = \pm 5$ 
r.l.u.~out of plane, and are offset vertically for clarity. The spectra shown 
in panel (a) were collected at $T = 1.7$ and in panel (b) at 40~K. Solid red 
lines show the total fit to the data, gray lines the background contribution, 
and filled areas the magnetic signal. We comment that some of the background 
curves in panels (a) include a weak inelastic peak arising from the twin 
contribution.}
\label{fig_multiple_fits2}
\end{figure}

\section{Determination of the spin Hamiltonian}
\label{sec:fit}

\begin{figure}[t]
\includegraphics[width=1\columnwidth]{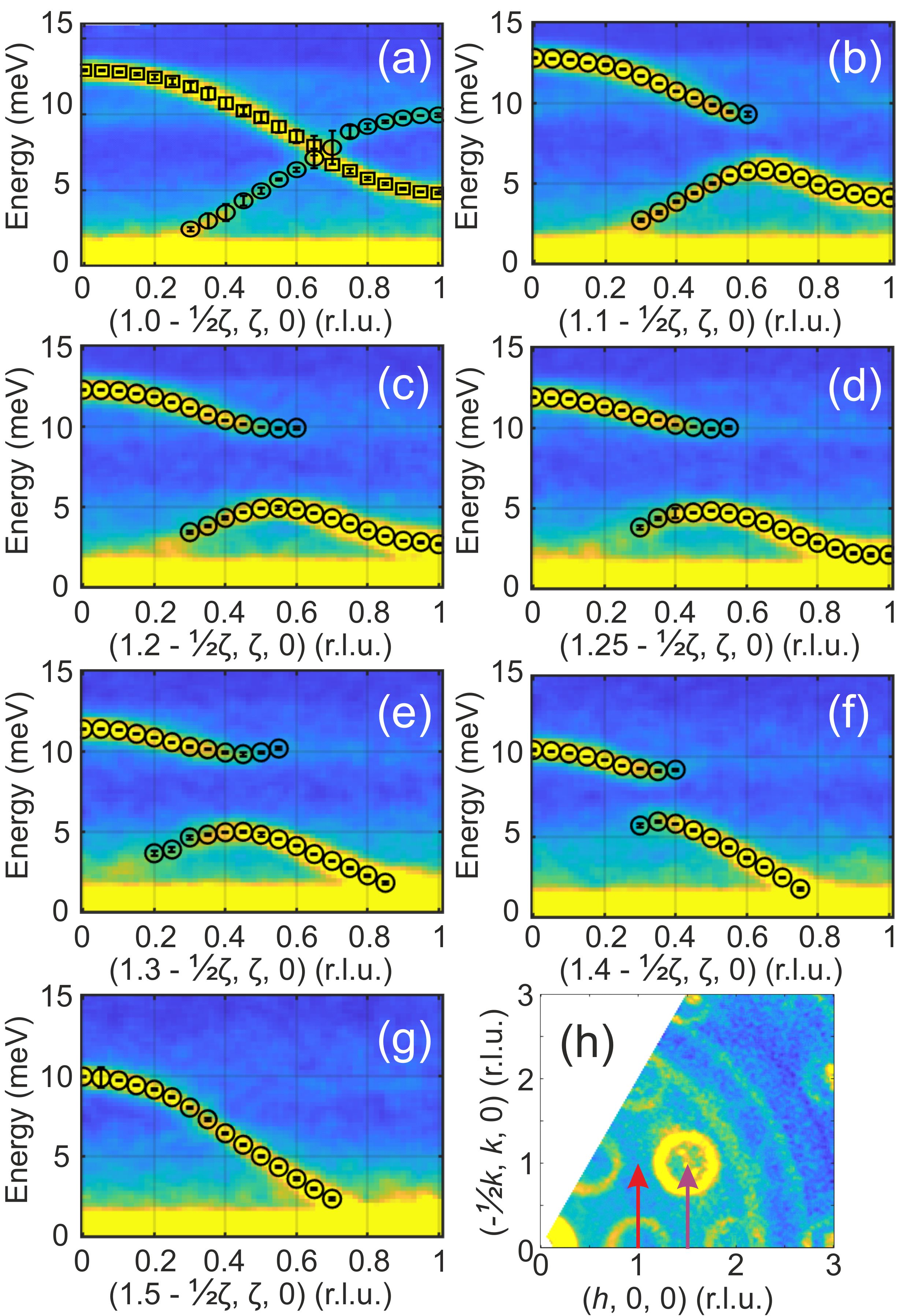}
\caption{(a-g) INS spectra of \crbr\ measured along multiple $\mathbf{Q}$ 
paths on PANTHER at $T = 1.7$~K with $E_{\mathrm{i}} = 30$~meV; the intensity 
scale is the same in all panels. The circles and error bars indicate 
respectively the mode positions and their statistical uncertainties 
obtained from fits to Eq.~(\ref{Eq:fit}).
(h)~Constant-energy cut in the $(h,k,0)$ scattering plane. Intensity data 
were obtained by integrating over the ranges $E = [2$--$3]$~meV and $l = 
\pm 5$ r.l.u. The red arrow shows the path used in panel (a), the purple 
arrow the path of panel (g), and panels (b-f) show parallel paths between 
the two arrows.}
\label{fig_multiple_fits}
\end{figure}

To extract the magnetic interaction parameters of \crbr\ we used both PANTHER 
datasets measured at $T = 1.7$~K, i.e.~with $E_{\mathrm{i}} = 15$ and 30~meV. 
We first quantified the positions of the magnon mode at 139 selected points 
in reciprocal space by fitting constant-$\mathbf{Q}$ cuts through the 
four-dimensional datasets to resolution-convolved Lorentz functions of the 
form  
\begin{align}
I(E) = a_0 + a_1 E^{-a_2} + \sum_{i = 1}^n \mathrm{exp} \bigg[ \frac{-E^2}
{2W_{\mathrm{res}}^2} \bigg] \! * \! \frac{I_i \, \Gamma_i^2 }{(E - \varepsilon_i)^2
 + \Gamma_i^2},
\label{Eq:fit}
\end{align}
where $a_0$, $a_1$, and $a_2$ are empirical background parameters, 
$W_{\mathrm{res}}$ combines the (Gaussian) instrumental resolution and 
integration-induced broadening, $I_i$, $\varepsilon_i$, and $\Gamma_i$ are 
the intensity, position, and width of magnon peak $i$, $n$ is the number of 
magnon peaks, and the * denotes convolution. Examples of this fitting 
procedure are shown in Fig.~\ref{fig_multiple_fits2}(a), where intensity data 
for 11 $\mathbf{Q}$ points on the path $(1.5~0~0) \rightarrow (1~0~0)$ 
exhibit a single, strong, resolution-limited magnon mode in each case, 
whose center position disperses with $\mathbf{Q}$. For comparison, in 
Fig.~\ref{fig_multiple_fits2}(b) we also present an analysis of the 
INS spectra collected at our highest measurement temperature, $T = 40$~K, 
where it is clear that the heights of the magnon peaks decrease considerably, 
they shift to lower energy, and they become significantly wider than the 
instrumental resolution, but remain clearly discernible. The positions 
and widths of the magnon modes extracted in this way were used in Fig.~3 
of the main text. Returning to base temperature, we used multiple series 
of fits of the type shown in Fig.~\ref{fig_multiple_fits2}(a) to obtain 
a set of points $\{ \varepsilon_i (\mathbf{q}) \}$, which together 
characterize fully the experimental dispersion. These extracted mode 
positions are shown on top of the corresponding $\mathbf{Q}$-$E$ cuts 
in Figs.~\ref{fig_multiple_fits}(a-g).

\begin{figure}[t]
\includegraphics[width=0.9\columnwidth]{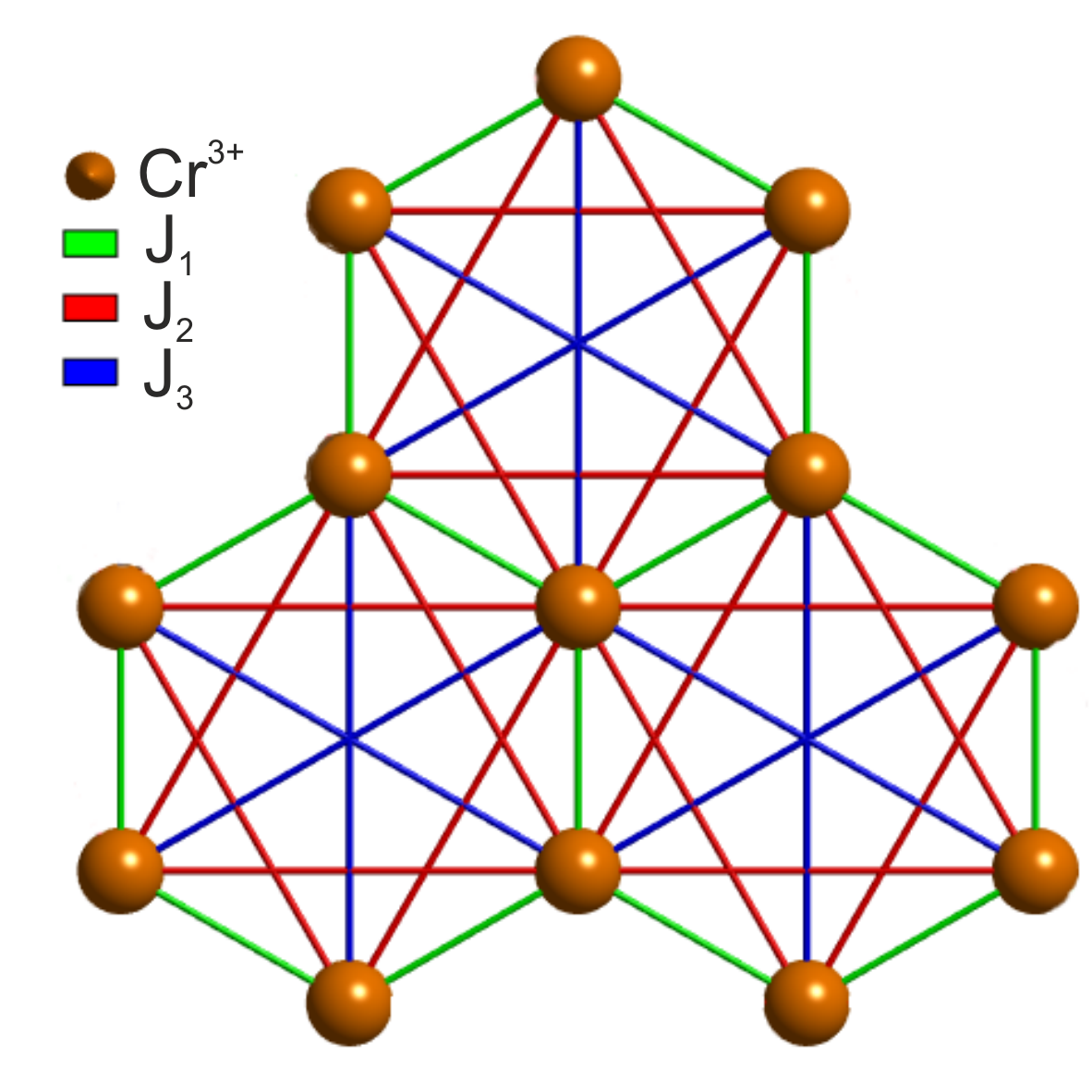}
\caption{Representation of magnetic interactions in the honeycomb lattice 
formed by the Cr$^{3+}$ ions (orange spheres) in a single layer of \crbr. 
Green, red, and blue lines indicate respectively the nearest- ($J_1$), second- 
($J_2$), and third-neighbor ($J_3$) Heisenberg superexchange interactions.}
\label{fig_ep}
\end{figure}

\begin{table}[b]
\caption{Magnetic interactions fitted to four spin models, which include 
different numbers of in-plane coupling parameters. $J_m$ and $D$ are 
respectively Heisenberg superexchange interactions and the single-ion 
anisotropy term, while $R_W$ quantifies the fit quality.} \label{tab:j}
\begin{ruledtabular}
\begin{tabular}{c| cccccc}
Model     & $J_1$   & $J_2$   & $J_3$   & $J_4$   & $D$    & $R_W$    \\
\hline
$J_1$         &  $-1.44$  & 0        & 0     & 0      &  $-0.2$   & 9.59   \\
$J_1$--$J_2$  &  $-1.46$  & $-0.057$ & 0     & 0      &  $-0.02$   & 5.81   \\
$J_1$--$J_3$  &  $-1.485$ & $-0.077$ & 0.068 & 0      &  $-0.028$  & 2.05  \\
$J_1$--$J_4$  &  $-1.494$ & $-0.068$ & 0.067 & 0.0082 &  $-0.03$   & 1.96   \\
\end{tabular}
\end{ruledtabular}
\end{table}

To deduce the spin Hamiltonian of \crbr, we used the \textsc{SpinW} package 
\cite{toth15} to calculate the low-temperature magnon dispersion and intensity 
of a model honeycomb FM, in order to compare the results with the set $\{ 
\varepsilon_i (\mathbf{q}) \}$ and with the corresponding measured intensities. 
As discussed in the main text, the models we consider have the form 
\begin{equation}
\mathcal{H} = J_1 \! \sum_{ i,j } \mathbf{S}_i \cdot \mathbf{S}_j 
 + J_2 \! \sum_{\langle i,j \rangle} \! \mathbf{S}_i \cdot \mathbf{S}_j 
 + J_3 \!\! \sum_{\langle\langle i,j \rangle\rangle} \!\! \mathbf{S}_i \cdot 
\mathbf{S}_j + D \sum_i (S_i^z)^2,
\label{Eq:Hamiltonianr}
\end{equation}
where the first, second, and third summations run over different nearest- and 
further-neighbor pairs of Cr$^{3+}$ ions, as indicated in Fig.~\ref{fig_ep}, 
and $D$ is the single-ion anisotropy term for the $S = 3/2$ spins. In fact 
we tested four spin models, taking into account in-plane superexchange 
interactions, $J_m$, up to fourth neighbors, with the results summarized in 
Table \ref{tab:j}. Starting with only a nearest-neighbor model, we found that 
introducing the second- and then the third-neighbor interactions each improves 
the quality of the fit quite considerably, with the optimal $J_2$ and $J_3$ 
values both being approximately 5\% of $J_1$. By contrast, the introduction 
of $J_4$ has only a very minor effect on the fit quality and returns a value 
one order of magnitude smaller than $J_2$ and $J_3$. We therefore conclude 
that the minimal model for an accurate description of the spin dynamics in 
\crbr\ is the $J_1$-$J_2$-$J_3$ model. We comment that, despite their small 
values, the effects of the $J_2$ and $J_3$ terms is in fact clearly visible 
in the magnon dispersions shown in Figs.~2 and 3 of the main text, because 
they act to lift the Dirac point from 6.5 meV, in a $J_1$ model that has 
two branches with energies 0 and 13 meV at the $\Gamma$ point, to 7.5 meV 
in \crbr, and a similar 1 meV shift is found at the M point. 

We comment that the resolution of our INS experiments did not allow us to 
determine precisely a possible spin gap at the $\Gamma$ point, where the 
signal is dominated by an elastic peak. However, a spin gap, $\Delta 
(\Gamma)$, can be measured very accurately by the method of ferromagnetic 
resonance (FMR), and for \crbr\ one finds $\Delta (\Gamma) = 0.08$~meV at 
$T = 2$~K \cite{dillon1962ferromagnetic,alyoshin1997rf}. With this 
information we performed our fits in two different ways: (i) considering 
only our INS data and (ii) taking into account the FMR value of $\Delta 
(\Gamma)$. Our fits (i) yielded a value $\Delta (\Gamma) = 0.2$ meV, which 
on the scale of the overall magnon band width is not a large discrepancy. 
All of the Heisenberg interactions remained the same within their error bars 
for both fitting procedures, with only the value of $D$ changing. Thus we 
fitted $D$ to the FMR value of $\Delta (\Gamma)$ and used the INS data without 
constraint to obtain the most accurate superexchange parameters for the spin 
Hamiltonian of \crbr. 

\section{Adapted interacting SWT calculation of magnon self-energies}
\label{sec:mod_LSWT}

To compare with our measurements of the thermal renormalization of the 
magnon dispersion and width, we followed Ref.~\cite{pershoguba2018dirac} 
to perform calculations of the magnon self-energy for the $J_1$-$J_2$-$J_3$ 
model of Sec.~\ref{sec:fit}. These authors computed the lowest-order 
spin-wave interaction terms, which are the Hartree contribution, $\Sigma_1 
(\mathbf{q})$, and the ``sunset'' contribution, $\Sigma_2 (\mathbf{q})$, 
for the situation with two magnon branches in the folded BZ that arises 
for the simplest non-Bravais lattices. In both contributions, one of the 
incoming magnons is thermally excited around the band minimum ($\mathbf{q}
 = 0$) and can be neglected in the interaction process, but causes the 
finite-temperature renormalization whose $T^2$ form is a straightforward 
consequence of the dimensionality factors summarized in the main text 
\cite{bloch30,pershoguba2018dirac}.

Considering first the Hartree part, $\Sigma_1 (\mathbf{q})$ is a real 
quantity that in the $J_1$ model is constant across the entire BZ as 
a consequence of the special property of these two bands. In the 
$J_1$-$J_2$-$J_3$ case, $\Sigma_1 (\mathbf{q})$ regains a weak 
$\mathbf{q}$-dependence due to the $J_2$ term. The 
$\mathbf{q}$-dependence of $\Sigma_2 (\mathbf{q})$ arises from an 
integral over one internal magnon momentum and takes the form 
\begin{equation}
{\rm Re} \, \Sigma_2 (\mathbf{q}) = \alpha T^2 \!\! \int \!\! d^2 p \, 
\frac{|v_{\mathbf{q};\mathbf{p}}|^2 \, (\varepsilon_{\mathbf{q}} - \varepsilon_{\mathbf{p}}
 - \varepsilon_{\mathbf{q} - \mathbf{p}})}{(\varepsilon_{\mathbf{q}}
 - \varepsilon_{\mathbf{p}} - \varepsilon_{\mathbf{q} - \mathbf{p}})^2 + \eta^2}
\label{eq:resigma}
\end{equation}
for the real part and 
\begin{equation}
{\rm Im} \, \Sigma_2 (\mathbf{q}) = \alpha \pi T^2 \!\! \int \!\! d^2 p \, 
|v_{\mathbf{q};\mathbf{p}}|^2 \, \delta (\varepsilon_{\mathbf{q}}
 - \varepsilon_{\mathbf{p}} - \varepsilon_{\mathbf{q} - \mathbf{p}})
\label{eq:imsigma}
\end{equation}
for the imaginary part. In these expressions, $\varepsilon_{\mathbf{q}}$ 
is the magnon dispersion relation, $\eta$ is a broadening function, and 
$|v_{\mathbf{q};\mathbf{p}}|$ denotes the matrix elements of the magnon-magnon 
interaction processes. The authors of Ref.~\cite{pershoguba2018dirac} made 
a detailed analysis of these matrix elements, but in the calculation of the 
real part encountered a historical problem \cite{dyson56a,dyson56b} arising 
in the long-wavelength limit and as a result reverted to some {\it ad hoc} 
matrix-element expressions. 

\begin{figure}[t]
\includegraphics[width=1\columnwidth]{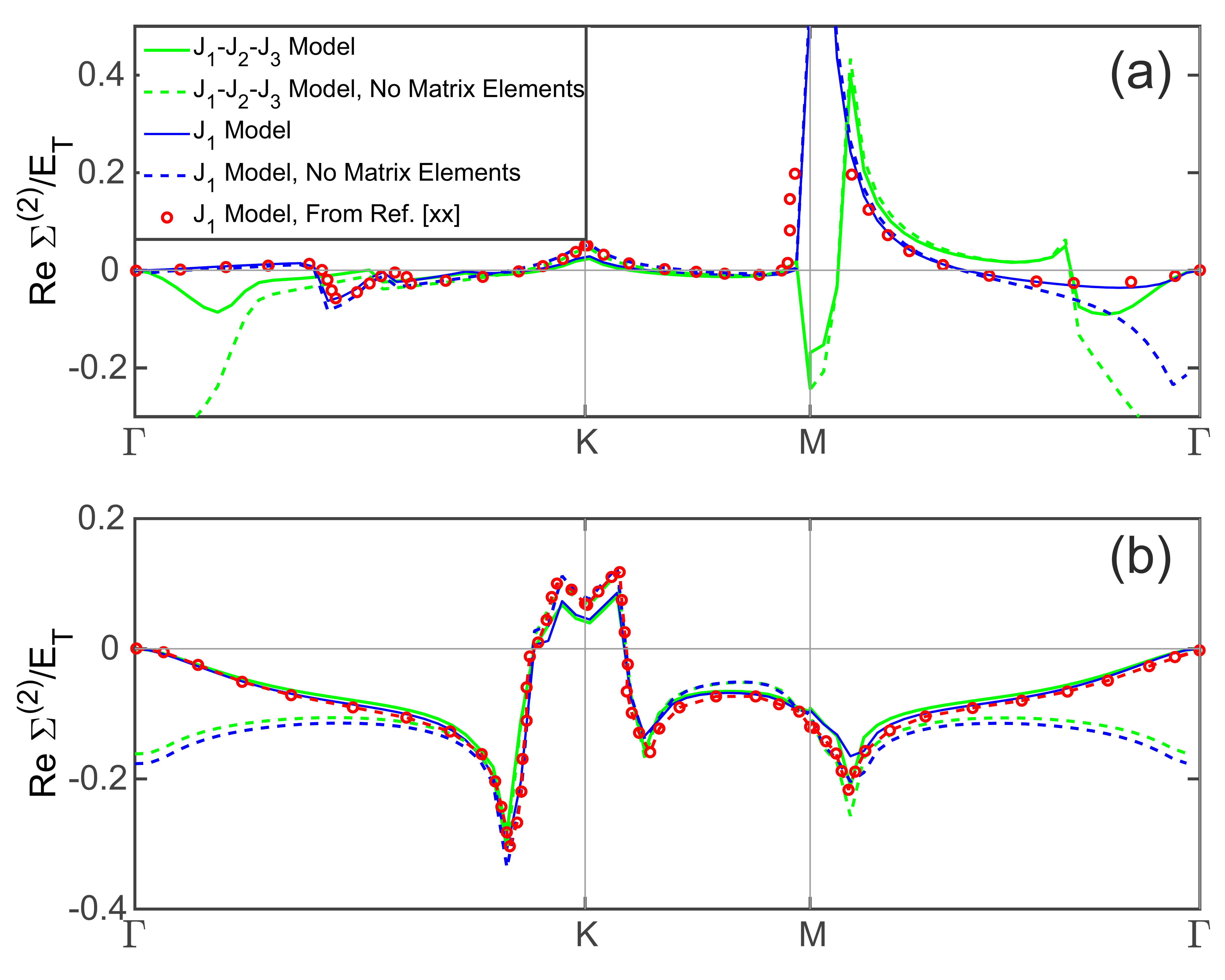}
\caption{Interacting SWT calculation of the real part of the magnon 
self-energy for the magnon branches above (a) and below (b) the Dirac-point 
energy. This band renormalization is computed for the nearest-neighbor ($J_1$) 
honeycomb model with and without the matrix elements proposed in 
Ref.~\cite{pershoguba2018dirac}, and is compared with the analogous results 
obtained for the $J_1$-$J_2$-$J_3$ magnon bands of Sec.~\ref{sec:fit}. The
unit of energy is $E_T = T^2/J_1 S^3$. }
\label{fig_resigma}
\end{figure}

\begin{figure}[t]
\includegraphics[width=1\columnwidth]{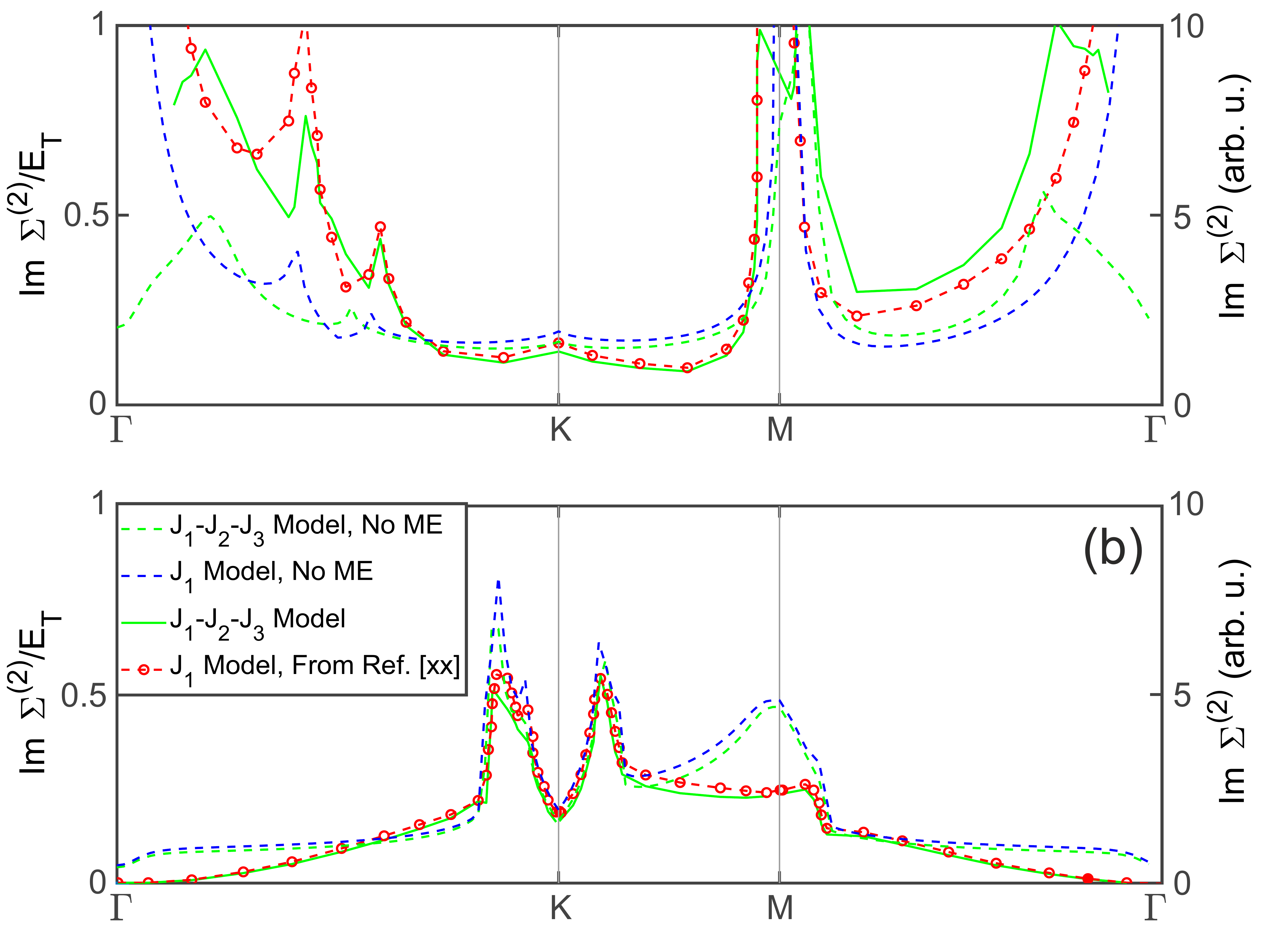}
\caption{Interacting SWT calculation of the imaginary part of the magnon 
self-energy for the magnon branches above (a) and below (b) the Dirac-point 
energy. The $\delta$-function part of the self-energy is computed for both 
the nearest-neighbor ($J_1$) and the $J_1$-$J_2$-$J_3$ magnon bands without 
including matrix elements (denoted ME), and then the magnon line width for 
the $J_1$-$J_2$-$J_3$ case is obtained by scaling the result of 
Ref.~\cite{pershoguba2018dirac} for the $J_1$ case with matrix elements. 
Calculations without matrix elements (dotted lines) are shown on the same 
arbitrary scale (right $y$ axis), while the results with matrix elements 
are shown in units of $E_{\mathrm{T}}$ (left $y$ axis).} 
\label{fig_imsigma}
\end{figure}

In computing the real and imaginary self-energies for the full $J_1$-$J_2$-$J_3$
magnon dispersion, we have followed Ref.~\cite{pershoguba2018dirac} by adopting 
their {\it ad hoc} matrix elements in Eq.~\eqref{eq:resigma} and by scaling 
their final result in Eq.~\eqref{eq:imsigma}. In this process we benefitted 
from their observation that the matrix elements are well behaved functions of 
wave vector that do not introduce any singular behavior and we performed 
calculations with $J_2 = J_3 = 0$ that reproduced theirs, as shown in 
Figs.~\ref{fig_resigma} and \ref{fig_imsigma}. This allowed us to obtain the 
final band renormalizations directly in Fig.~\ref{fig_resigma}, where the 
structure of the calculation requires attributing the self-energies to the 
upper and lower magnon branches, rather than to modes 1 and 2, and to obtain 
the final line widths in Fig.~\ref{fig_imsigma} by scaling to our results with 
the experimental values of $J_2$ and $J_3$. We also benefitted from the analysis
of prefactors ($\alpha$ in Eq.~\eqref{eq:resigma} and \eqref{eq:imsigma}) 
provided in Ref.~\cite{pershoguba2018dirac} such that our calculated  
magnitudes of the magnon shift and line width have the same ``parameter-free'' 
status as theirs. In relating the calculated ${\rm Im} \, \Sigma (\mathbf{q})$ 
to the measured line widths in Fig.~3(e) of the main text, we assumed that 
they correspond to the HWHM of the Lorentzian fitting function 
[Eq.~\eqref{Eq:fit}].

By inspecting the $\mathbf{q}$-dependence of the thermal renormalization 
and line width, we find that the interacting SWT calculations for the 
$J_1$-$J_2$-$J_3$ bands continue to predict multiple characteristic peaks. 
With the exception of divergences in the line width at $\Gamma$ and M, which 
are consequences of perfect nesting in the $J_1$ model, all of the van Hove 
cusps found in Ref.~\cite{pershoguba2018dirac} are only slightly moved or 
blunted. By contrast, our reduced data in Figs.~3(d) and 3(e) of the main text 
show a total lack of characteristic features as a function of $\mathbf{q}$. 
Beyond the flat response around the M point shown in Fig.~4 of the main 
text, we also do not find evidence, within the uncertainties of our data, 
for the peak near the K point highlighted in Ref.~\cite{pershoguba2018dirac} 
(we comment that the appearance of this feature in the results of 
Ref.~\cite{yelon1971renormalization} is in fact based on a single data 
point). We remark that the largest values of the reduced band shift and 
line width visible in Figs.~3(d) and 3(e) of the main text appear where the 
magnon energy in the denominator of Eq.~(2) of the main text is small, and 
this is the reason for the large uncertainties on all of these data points. 

Thus one must conclude that the van Hove features are artifacts of the low 
level of approximation in the interacting spin-wave calculation. In particular, 
the energetic renormalization of the band, ${\rm Re} \, \Sigma (\mathbf{q})$, 
can be expected to have a significant effect on the nesting contributions to the
interaction terms, and it would be necessary to include this self-consistently. 
It is also possible that the $S = 3/2$ nature of the quantum spins should be 
included in a constrained spin-wave treatment, and that it may have significant 
consequences for nesting effects even at the $1/S$ level. Beyond a more 
sophisticated spin-wave treatment, in the main text we also draw attention 
to the rapid advances taking place in the numerical calculation of spectral 
functions for quantum magnetic models within the framework of stochastic 
analytic continuation quantum Monte Carlo methods and separately within 
the framework of tensor-network techniques.

\end{document}